\documentclass[12pt]{article}
\pdfoutput=1
\usepackage{epsfig}
\usepackage{amssymb,amsmath}
\usepackage{physics}
\usepackage[utf8]{inputenc}
\usepackage{epsfig}
\usepackage{amsfonts}
\usepackage{amssymb}
\usepackage{mathtools}
\usepackage{amsmath}
\usepackage{cancel}
\usepackage{color}
\usepackage{slashed}
\usepackage{cite}
\usepackage{calligra}
\usepackage{caption}
\usepackage{subcaption}
\usepackage{comment}
\usepackage{marginnote}
\usepackage{accents}
\usepackage{url}
\usepackage{array,graphicx,caption}

\allowdisplaybreaks

\begin{document}

\begin{center}

\vspace{1cm}

\renewcommand{\thefootnote}{\fnsymbol{footnote}}

{\bf \Large Scattering matrix for current-current 
deformations of the (2,2s+1) RSOS models} \vspace{1.0cm}

{\large Y. Pugai$^{1,2}$, D. Shabetnik$^{1,3}$}
\vspace{0.5cm}

{\it
$^1$Kharkevich Institute for Information Transition Problems,\\ Russian Academy of Sciences,\\  127994, Moscow, Russia\\
$^2$Landau Institute for Theoretical Physics,\\ Russian Academy of Sciences,\\ 142432, Chernogolovka, Russia\\
$^3$Moscow Institute for Physics and Technology, \\ 
141700, Dolgoprudny, Russia\\

E-Mail: ~\rm{slava@itp.ac.ru},~\rm{shabetnik.da@phystech.edu}
}

\renewcommand{\thefootnote}{\arabic{footnote}}
\setcounter{footnote}{0}

\vspace{1cm}

\abstract{We study (2,2s+1) RSOS lattice models deformed by the current-current operator. Solving the deformed Bethe ansatz equations for the model in the regime III we find explicit expressions for the ground state energy as well as for the energy, the momentum and the scattering matrix of the breather-like excitations. In the scaling limit we  get an additional confirmation to the proposal that the bi-local deformations are a lattice analogue of the $T\bar{T}$ perturbations of integrable massive QFTs proposed by Smirnov and Zamolodchikov.}
\end{center}


\newpage
\newpage
\section{Introduction}
The integrability of two-dimensional lattice models and closely related massive Quantum Field theories is based on the existence of commuting integrals of motion. In both classes of exactly solvable models the central role is played by the Yang-Baxter equation. For lattice models this is a functional equation for Boltzmann weights or R matrices \cite{Baxter:book}; in this case row-to-row (or diagonal, or corner) transfer matrices commute. Due to integrability 
one can find physically important observables, such as free energy, form factors and correlation functions. In particular, it is possible to derive the scattering matrix of the low-laying excitations; the S-matrix also satisfies the Yang-Baxter equation. 

In the scaling region universal properties of a lattice model  are described by a corresponding Quantum Field Theory. The basic ideas about integrable field theories were maintained in the pioneer's work of Zamolodchikov and Zamolodchikov  \cite{ZZ} where the exact S-matrix for solitons in the quantum sine-Gordon model was found. Due to the existence of an infinite set of integrals of motion the scattering is factorized and it is possible to find the exact S-matrices as a solution of the Yang- Baxter equation and the unitarity, crossing and bootstrap closing conditions. The scattering data fixes the theory. Starting from an exact pure elastic S matrix it is possible to apply different powerful methods to analyse observables for the corresponding QFT in the bootsrtap approach.

As mathematical objects, both R- and S-matrices can be equally treated as solutions of the YBE and other equations like unitarity and crossing. However there is a subtle difference between them. Indeed, YBE in general is a matrix equation. Therefore it is possible to generate its new solutions by multiplying the known ones to a scalar factor with good analytic properties. For integrable Boltzmann weights this operation does not change the theory. It is equivalent to a simple rescaling of the normalization of the Boltzmann weights. This redefinition is not important for the correlation functions, for example. As for the scattering theory, however, multiplying a solution of the triangle equation to a scalar (CDD) factor which does not destroy unitarity, crossing and bootstrap fusion equations lead, in general, to a different scattering data, i.e. to another field theory. It was not clear in earlier years of integrable QFT what is the meaning of this freedom - to deform scattering matrices by CDD multiples.

In ref. \cite {TT} Smirnov and Zamolodchikov proposed a general construction  of the deformation of massive integrable QFTs by the  special operator of the form $T\bar{T}$ \cite{ZamTT} which leads to the theories with  CDD deformed S-matrices. (For earlier results in this field see, for example,  refs. \cite{AlZamTri, AlZamStair,Dorey,MussardoSimon}).  
In our work we would like to understand some details of this construction using lattice regularization which can be simpler than a QFT. More explicitly, we want to re-check the correspondence between two classes of integrable deformations: the bi-local deformations of lattice models by Bargheer, Beisert, and Loebbert \cite{LR} and  the $T\bar{T}$ deformations of an integrable QFT provided by  Smirnov and Zamolodchikov \cite {TT}. The statement on the correspondence of the two classes of integrable perturbations was already provided in refs. \cite{TTspin, TTspin1}. Both constructions of integrable deformations are general and can be applied to a broad set of integrable models. We restrict our attention to the simplest theory with breather-like excitations. 

Namely, the main subject of our studying is the two dimensional integrable lattice RSOS  model of Forrester-Baxter with the corresponding parameters $(2,2s+1)$  \cite {Forrester-Baxter}. In the regime III the critical behaviors of these  models are described by  minimal models $\mathcal{M}_{2,2s+1}$ of CFT \cite{BPZ} with the central charge of the Virasoro algebra 
$$
c=1-\frac{6}{\xi(\xi+1)}\,,
$$
where $\xi=2/(2s-1)$. In the scaling region in a vicinity of the critical point the theory can be described as the perturbation of the CFT by the energy operator $\Phi_{13}$
$$
{\mathcal A}_{\xi}=\mathcal{M}_{\xi}+\lambda \int d^2x \Phi_{13}\,.
$$
The operator $\Phi_{13}$ is relevant and the coupling constant is proportional to $m^{4/(\xi+1)}$. The integrability of the massive $\Phi_{13}$ perturbation described above follows from the fact that an infinite number of the commuting integrals of motion survive \cite{Zam}. 

Any massive integrable QFT can be equivalently treated as a factorized scattering theory. In our case the spectrum consists of $s-1$ breathers with the diagonal S-matrix which is known from the quantum sine-Gordon model \cite{ZZ}. More explicitly, we consider the so-called Restricted sine-Gordon model \cite{AlZam,LeClair,SmirnovRG,Freund}.

Our initial motivation was related to the fact that  we could not find a direct derivation of the scattering matrix for the low-energy excitations in the off-critical lattice Forrester-Baxter model in the regime III. In the unitary case many exact results are known \cite{BR}; several results are also available for the eight vertex model with the appropriate crossing parameter. Trying to naively extend these results from the unitary case to non-unitary, we realised that the vacuum structure as well as details of the Bethe ansatz string solutions strongly depend on the arithmetic properties of the parameter of the model $\xi$, so both cases should be considered independently. Our initial aim was to fill this gap: to study  vacuum and excited states and find elliptic scattering matrices of the  Forrester-Baxter lattice models in the regime III. 

As we already mentioned, another motivation of the studying is related to our attempts to understand the results of ref. \cite{TT}. In our case  new models  can be defined as the $T\bar{T}$ flow of the action
	\begin{equation}
\frac{d}{d\alpha}	\left({\mathcal A}_{\xi}\right)^{(\alpha)}=\int \left(T\bar{T}\right)^\alpha d^2x\,,
	\label{ZS}
	\end{equation}
where $\alpha$ is the deformation parameter and operator $(T\bar{T})^{(\alpha)}$ is constructed from the components of the conserved stress-energy tensor of the deformed theory. 

Since the perturbation is irrelevant, there are some unusual phenomena in Wilson's approach \cite{Wilson} to QFT: the renormalization procedure demands, in general, an infinite number of counterterms; the $T\bar{T}$ flow is directed toward the critical point, while the $\Phi_{13}$ flow is going away from it, etc. \cite{AlZamTri,AlZamStair,Dorey,MussardoSimon}. However, the fact that the perturbed theory remains integrable drastically simplifies the model and many interesting physical quantities still can be computed exactly.

One of the remarkable features of the perturbation operator $(T\bar{T})^{(\alpha)}$ is the factorization property \cite{ZamTT}. It allows to derive a differential equation, known as the inviscid Burgers' equation, for the energy spectrum $E^{(\alpha)}$ of the deformed model \cite{TT}
\begin{equation*}
  \frac{\partial}{\partial{\alpha}}E^{(\alpha)}+E^{(\alpha)}\frac{\partial}{\partial{L}}E^{(\alpha)}+\frac{P^2}{L}=0.
  \end{equation*}
Here $P$ is a total momentum of an eigenstate and $L$ is the size of the system. The equation can be solved exactly by the method of characteristics. 
The solution of this equation for the ground state  $(P=0)$ reads
\begin{equation*}
   \frac{1}{L}E^{(\alpha)}=\frac{E_0}{1+\alpha E_0}\,,
\end{equation*}
where $E_0$ is the energy density of the undeformed ground state \cite{AlZam,AlZamTBA}. In general, the eigenvalues of integrals of motion also start to depend on the deformation parameter $\alpha$. For example, the masses of the particles $m^{(\alpha)}$ in the corresponding scattering theory are changed in comparison with the pure case $m=m^{(0)}$ 
\begin{equation*}
   m^{(\alpha)}=\frac{m}{1+\alpha E_0}\,.
\end{equation*}

Since the model is integrable, the S-matrix can be calculated exactly. The proposal is that the S-matrix of the deformed theory differs from the undeformed one by the CDD multiple  \cite{TT, MussardoSimon} (see also \cite{CNST, DeformedS1,DeformedS2,DeformedS3, DeformedS35, CardyTT, DeformedS4}). In ref. \cite{TT} a generalization of this statement to the case of currents of higher spins was also proposed.  

For theories (\ref{ZS}) this statement was confirmed by the computing of the first order corrections in the coupling constant $\alpha$ to the S-matrix \cite{ZZ} using the four particle form factors \cite{Smirnov} of the $T\bar{T}$ operator and it's higher currents analogues \cite{JMS, DelfinoNikoli1,DelfinoNikoli2, DelfinoNikoli3, TTSinh,TTSinh1} in the breather sector of the sine-Gordon model. There are also heuristic arguments for deriving the CDD factor expression which are based on the Burgers' equation \cite{TT}. However, we could not find a direct derivation of the CDD multiple in the breather-like S-matrix in all orders of the perturbation theory. 

As it was mentioned above, the perturbation by the $T\bar{T}$ operator is irrelevant one and the field theory is rather complicated. Usually, dealing with QFT it is useful to have a convenient regularization procedure at small distances. So it is reasonable to look for the lattice analogues of the models (\ref{ZS}). Among an infinite set of lattice models having the same scaling limit the most simple are those which do not destroy the integrability. The last one is an essential element of a beauty and a simplicity of the $T\bar{T}$ deformations. It was observed in refs. \cite{TTspin, TTspin1} that both $T\bar{T}$ and lattice bi-local deformations proposed by Bargheer, Beisert, and Loebbert in ref. \cite{LR} share the same algebraic features. Namely, it was derived that the expectation value of the product of conserved currents of the XXZ model factorizes in the same way as it is was proposed for QFTs \cite{ZamTT}. Due to this phenomena, it is expected that the deformation by the operator constructed from the lattice energy and momentum also leads to the appearance of a CDD-like factor in the corresponding asymptotic (magnon) S-matrix \cite{LR, LR1,LR2}:
\begin{eqnarray}
S^{(\kappa)}(u,v)=S(u,v)\exp\Big({i\kappa(e(v)p(u)-e(u)p(v))}\Big)\,,
\label{MagS1}
\end{eqnarray}
where $\kappa$ is the deformation parameter for the lattice model. It is reasonable to assume that $\kappa$ in a lattice model has a similar meaning as $\alpha$ in a QFT. 

Strictly speaking, the multiple in the S-matrix (\ref{MagS1}) is not a CDD-factor. However this extra multiple is very special: the function in the exponent is a quadratic combination of the Hamiltonian and momentum eigenvalues $e(u), p(u)$ under the action at the one-magnon state. In the QFT limit this factor is expected to become a genuine CDD multiple.

We would like to check if the CDD-like deformation of the magnon S-matrix (\ref{MagS1}) indeed  leads to the the same deformation of the breather S-matrix in the scaling limit. The arguments on correctness of the hypothesis \cite{TTspin, TTspin1} are based on the idea that the construction of the ref. \cite{LR} is algebraic. Therefore it is rather natural to expect that the structure of final answers should not depend on a particular regime. Despite the fact that one-particle eigenvalues are different in different regimes, the CDD-like multiples in the corresponding scattering matrices should have the same structure: it should be the combination of the eigenvalues of the momentum and energy of low-laying excitations. However, it was not obvious for us how these factors appear in the breather S-matrix if one starts from the deformed Bethe ansatz equations without using additional algebraic arguments. One can expect that there should be a subtle mechanism related to a very special nature of the deformation due to which the correct QFT answers \cite{TT}  appear in the breather sector in this approach. Our main aim is to check details of this mechanism accurately.

This work is organized as follows.
In the section 2 we recall basic notions and definitions for the Forrester-Baxter models $(2,2s+1)$ and their bi-local current-current deformations along the lines of the construction \cite{LR}.

In section 3 we study the deformed Bethe ansatz equations for the ground state and for string excitations. It turned out that the corresponding linear integral equations can be solved exactly. Using the solutions, we derive the energy density of the deformed ground state
\begin{equation}
    E_{g}^{(\kappa)}=\frac{\mathcal{H}}{1-\kappa \mathcal{H}}\,,
 \end{equation}
where $\mathcal{H}$ is the density of the ground state energy in the pure case $(\kappa=0)$. This result is consistent with the solution of the Burgers' equation discussed above if $P=0$ and $\alpha=-\kappa$. Also we find the 
energies $\mathcal{E}_a^{(\kappa)}(u)$ and momenta $\mathcal{P}_a^{(\kappa)}(u)$ of  excitations
\begin{eqnarray}
&&
\mathcal{E}_a^{(\kappa)}(u)=\frac{\mathcal{E}_a(u)}{1-\kappa \mathcal{H}}\,,\\
&&
\mathcal{P}_a^{(\kappa)}(u)=\frac{\mathcal{P}_a(u)}{1-\kappa \mathcal{H}}\,,
\end{eqnarray}
where the index $a$ denotes the length of the corresponding string solution of the Bethe ansatz equations. The corresponding energies and momenta of the particles, respectively $\mathcal{E}_a(u)$ and $\mathcal{P}_a(u)$,  in the undeformed case have the following form: 
\begin{equation}
    \mathcal{P}_a(u)=-\frac{1}{i}\ln{\frac{\Theta_2\Big( \frac{\pi}{2}(u+\frac{a\xi}{2}) |\frac{i\pi}{2\epsilon}\Big)\Theta_1\Big( \frac{\pi}{2}(u-\frac{a\xi}{2}) |\frac{i\pi}{2\epsilon}\Big)}{\Theta_1\Big(\frac{\pi}{2}(u+\frac{a\xi}{2})  |\frac{i\pi}{2\epsilon}\Big)\Theta_2\Big(\frac{\pi}{2}(u-\frac{a\xi}{2})|\frac{i\pi}{2\epsilon}  \Big)}},
\end{equation}
\begin{equation}
    \mathcal{E}_a(u)=-\frac{ d } { du }\ln{\frac{\Theta_2\Big( \frac{\pi}{2}(u+\frac{a\xi}{2})|\frac{i\pi}{2\epsilon} \Big)\Theta_1\Big( \frac{\pi}{2}(u-\frac{a\xi}{2})|\frac{i\pi}{2\epsilon} \Big)}{\Theta_1\Big(\frac{\pi}{2}(u+\frac{a\xi}{2}) |\frac{i\pi}{2\epsilon} \Big)\Theta_2\Big(\frac{\pi}{2}(u-\frac{a\xi}{2})  |\frac{i\pi}{2\epsilon}\Big)}}.
\end{equation}
Using these results we find  in section 4 that the deformed S-matrix of elliptic breathers is
\begin{equation}
    \mathcal{S}^{(\kappa)}_{a,b}(u,v)=\mathcal{S}_{a,b}(u,v)e^{\frac{ i\kappa}{1-\kappa\mathcal{H}}\big(\mathcal{P}_{a}(u)\mathcal{E}_{b}(v)-\mathcal{E}_{a}(u)\mathcal{P}_{b}(v)  \big)},
\end{equation}
where ${S}_{a, b}(u, v)$ is the exact scattering matrix in the pure case in the regime III of the Forrester-Baxter model:
\begin{equation}
\mathcal{S}_{a, b}(u, v)= \prod\limits_{l_1=1}^{a}\prod\limits_{l_2=1}^{b}\mathcal{S}_{1,1}\Big(u-v-\frac{\xi}{2}(a-2l_1-b+2l_2)\Big).
\end{equation}
This result is  in agreement with the fusion structure of the corresponding restricted sine-Gordon model; the basic scattering matrix of elliptic breathers is  
\begin{equation}
\mathcal{S}_{1,1}(u)=\frac{\Theta_1(\pi\frac{u+\xi}{2}|\frac{i\pi}{2\epsilon})\Theta_2(\pi\frac{u-\xi}{2}|\frac{i\pi}{2\epsilon})}{\Theta_1(\pi\frac{u-\xi}{2}|\frac{i\pi}{2\epsilon})\Theta_2(\pi\frac{u+\xi}{2}|\frac{i\pi}{2\epsilon})}.
\end{equation}
We see that the bootstrap equations are automatically satisfied for the deformed S-matrix as well. There are no extra poles in the physical strip and 'unitarity' and crossing relations are satisfied. These are the basic features of the CDD multiple. On the lattice, however, the  relativistic symmetry is broken and the scattering matrix depends  on both rapidities and not on their difference. 

Next we consider a scaling limit of the RSOS models and find that the corresponding lattice deformations are in agreement with the $T\bar{T}$ deformed restricted sine-Gordon model. The subtle point is that we have to normalize the vacuum expectation value of the energy field to be zero in the conformal limit. This leads to the necessity to modify a deformation of the Bethe ansatz equations. Then we are able to reproduce answers of Smirnov and Zamolodchikov with the correspondence $\alpha=-\kappa$, ${\mathcal H}\to E_0$.  

In particular we find that in the scaling limit the scattering matrix of low-laying excitations become breather S-matrix of the restricted sine-Gordon model  deformed by a genuine CDD factor \cite{ZZ}
\begin{equation}
    \mathcal{S}^{(\alpha)}_{a,b}(\beta)=\mathcal{S}_{a,b}(\beta)e^{\frac{i\alpha}{1+\alpha E_0}m_a m_b\sinh{\beta}}\,.
\end{equation}
We also provide heuristic arguments that this structure can be expected. 

A short summary of the results on the correspondence between current-current perturbations of the RSOS model in the scaling limit and the $T\bar{T}$ perturbations of the Restricted sine-Gordon model is given in the section 5. 

\section{Forrester-Baxter models and their deformations}
The lattice models studied in this work are the Forrester-Baxter models of the type $(2,2s+1)$  introduced in ref. \cite{Forrester-Baxter}. These RSOS models are non-unitary generalizations of the well-known Andrews, Baxter and Forrester models \cite{ABF:1984}. 

Consider a square lattice of the size $N\times N$. In what follows we assume that $\frac{N}{2}$ is an integer. The fluctuating variables $l$, called heights, are associated with the sites of the lattice. These variables are integers and their admissible values are restricted:
$
    1\leq l\leq 2s.
$
According to the SOS prescription the heights in the nearest neighbouring sites differ by $\pm 1$.

\subsection{Boltzmann weights of RSOS model}
The local Boltzmann weight is assigned to a configuration of four heights round a face of a lattice as it is shown in the figure.
\begin{figure}[!ht]
\centering
\includegraphics[width=0.4\linewidth]{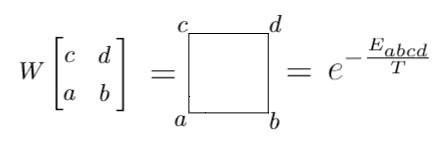}
\end{figure}
Here $E_{abcd}$ is the energy prescribed to the face. The integrability of the model follows from the fact that the local Boltzmann weights satisfy the Yang-Baxter equation \cite{Baxter:book}. The explicit parameterization of the Boltzmann weights can be given in terms of the function of the spectral parameter $u$
\begin{equation}
    [u]=\Theta_1\left(\frac{\pi u}{\xi+1}\biggl|\frac{i\pi}{(\xi+1)\epsilon}\right).
\end{equation}
Here $\epsilon$ is a temperature-related parameter which measures the derivation from the criticality and $\xi$ is the parameter of the model:
\begin{eqnarray}
\xi=\frac{2}{2s-1}\,.
\label{XiParameter}\end{eqnarray}
Here $\Theta_1(u|\tau)$ is a Jacobi elliptic theta function. We use the following standard representation: 
\begin{equation*}
    \Theta_1(u|\tau)=2q^\frac{1}{4}\sin{u}\prod\limits_{n=1}^{\infty}(1-q^{2n})(1-2q^{2n}\cos{u}+q^{4n}),
\end{equation*}
where the elliptic nome is $q=e^{i\pi\tau}$. 
Note that by the definition the function $[u]$ is periodic with the real period $2\xi+2$ and the imaginary quasi-period ${i\pi}/{\epsilon}$. 

Due to the RSOS restrictions of the model only specific configurations are allowed also in the integrable case the Boltzmann weights should satisfy the Yang-Baxter equation in the IRF from.  The explicit form of non-trivial weights in the Forrester-Baxter model is given as \cite{Forrester-Baxter}
\begin{eqnarray}
&&
W\left[\begin{matrix}
 l \pm1 & \l\\
l &  l\mp1
\end{matrix}\biggl| u\right]=\frac {[1-u]} {[1]},\cr
&&
W\left[\begin{matrix}
 l & l\pm1\\
 l\mp1 & l
\end{matrix}\biggl| u\right]=-\frac {[l\pm1]} {[l]} \frac {[u]} {[1]},\cr
&&
W\left[\begin{matrix}
 l & l\pm1\\
l\pm1 & l
\end{matrix}\biggl| u\right]=\frac {[l \pm u]} {[l]}\,.
\label{BoltzmanWeight}
\end{eqnarray}
Note here, that there is a freedom in changing the overall normalization of the weights.

There are several different regimes in the theory. We restrict attention to the so-called regime $III$ where $0<u<1$ and  $0<\epsilon<\infty$. If $\epsilon \to \infty$, the system is 'frozen' in it's unique ground state. The last one consists of heights which are equal to either $s$ or $s+1$ at SW-NE direction while along the SE-NW direction the heights are alternating. This simple structure of the ground state is specific for the $(2,2s+1)$ Forrester-Baxter models. In more general cases there are several ground state configurations and their structure depends on the details of the continuous fraction of the number $\xi$ 
\cite{Takahashi_strings,Foda,Foda1}. We start from the system of finite size with the periodic boundary conditions. Respectively, in the thermodynamic limit we assume that heights at the sites which are situated far away from the 'center' of the lattice coincide with those in the ground state configuration. 

At the critical point ($\epsilon\rightarrow 0$) the Forrester-Baxter lattice models undergo a second order phase transition. In the criticality the models become scaling and conformally invariant. They are described by the minimal models $M_{2,2s+1}$ of Belavin, Polyakov and Zamolodchikov with the central charge of the corresponding Virasoro algebra
\begin{equation*}
    c=1-\frac{3(2s-1)^2}{2s+1}\,.
\end{equation*}
For example, for the $s=2$ case we get the conformal Lee-Yang model $\mathcal{M}_{2,5}$ with the central charge of the Virasoro algebra $c=-22/5$. 

In the vicinity of the critical point the corresponding scaling model can be treated as a massive integrable QFT known as the restricted sine-Gordon model. The  scaling model can be considered as a perturbation of the conformal field theory by the relevant energy operator \cite{Zam}. For example, the $(2,5)$ RSOS model in the scaling limit turns out to be the massive Lee-Yang model \cite{Cardy,AlZam}. Equivalently the integrable QFTs can be described as the factorized scattering theories. For the scaling $(2,2s+1)$ Forrester-Baxter model the corresponding factorized scattering theory consists of $s$ breathers with the diagonal two-particle scattering matrix \cite{Freund,SmirnovRG,LeClair}. One of our hypothesises is that in the elliptic case the low lying excitations also have  breather-like structure. The argument in favour to this comes from the fact that the algebraic approach to the correlation functions \cite{LHP} and form-factors \cite{SL,YP,FR, LPST} is consistent with that proposal.

\subsection{Transfer matrix and integrals of motion}
We want to study  integrable deformations of the Forrester-Baxter models in the spirit of the approach developed in ref. \cite{LR}. To introduce these deformations we need to recall some basic concepts and definitions. 

We define the row-to-row transfer matrix $\mathbf{T}_{\vec{a}}^{\vec{b}}(u)$ as the product of all Boltzmann weights in a row of the lattice
\begin{eqnarray*}
\mathbf{T}^{\vec{b}}_{\vec{a}}(u)=\prod\limits_{j = 1}^{N}W
\left[\begin{matrix}
b_j & b_{j+1}\\
a_j &  a_{j+1}
\end{matrix}\biggl| u\right],
\end{eqnarray*}
where $\vec{a}=(a_1,..,a_N)$, $\vec{b}=(b_1,..,b_{N})$ are the heights at the sites in the given row and the periodic boundary conditions are assumed
\begin{eqnarray*}
&&b_{N+1}=b_1, \cr && a_{N+1}=a_1.
\label{PBC}
\end{eqnarray*} 
Since the system is integrable, the transfer matrices with different spectral parameter commute 
\begin{eqnarray}
[\mathbf{T}(u), \mathbf{T}(v)]=0\,.
\end{eqnarray}
This implies that there exists the infinite number of commuting integrals of motion - the conserved charges $\mathbf{Q_r}$   ($r=1,2..$). They can be defined by expanding the logarithm of the transfer matrix into a series in terms of the spectral parameter
\begin{equation}
  \mathbf{Q}_r=\frac{d^r}{du^r}\ln{\mathbf{T}(u)}\biggr|_{u=0}=\sum_j \mathbf{q}_r[j],
  \label{IntQ}\end{equation}
where we pointed out that $\mathbf{Q}_r$ is an extensive operator: it can be represented as a sum over local operators which act non-trivially only on a finite number of heights residing in neighboring faces of the lattice. 

The $r=0$ term in the expansion of the transfer matrix corresponds to the shift operator. Using the explicit form of the Boltzmann weights, we find that  
\begin{eqnarray*}
\exp\left(i \mathbf{P}\right)_{\vec{a}}^{\vec{b}}=\mathbf{T}(0)_{\vec{a}}^{\vec{b}}=\prod_{j}\delta^{b_{j+1}}_{a_j}\,,
\end{eqnarray*}
where $\mathbf{P}$ is the lattice momentum operator. This operator is exceptional in the sense that it can not be represented as a sum over local operators in comparison with (\ref{IntQ}).  The $r=1$ integral of motion is the Hamiltonian
\begin{eqnarray}
\mathbf{H}=\frac{d}{du}\ln{\mathbf{T}(u)}\biggr|_{u=0}=\sum_j {\mathbf{q}_1}[j]\,,
\label{hamiltonian}
\end{eqnarray}
where the Hamiltonian density $\mathbf{q}_1[j]$ can be expressed in terms of the Boltzmann weights as follows
\begin{eqnarray}
\left(\mathbf{q}_1[j]\right)_{\vec{a}}^{\vec{b}}=\Bigg(\prod_{i<j} \delta^{b_i}_{a_i}\Bigg) \ 
\frac{d}{du}W\left[\begin{matrix}
 a_{j-1} & b_j\\
 a_j & a_{j+1}
\end{matrix}\biggl| u\right]\biggr|_{u=0}\ 
\Bigg(\prod_{i>j} \delta^{b_i}_{a_i}\Bigg).
\end{eqnarray}
We see that this local operator acts non-trivially only on three adjacent sites $a_{j-1},a_{j},a_{j+1}$ while it acts as the identity operator on the rest of cites. Higher integrals of motion, which correspond to $r\geq 2$,  can be obtained in the similar way.  
For example, the density of the $r=2$ charge is 
\begin{eqnarray*}
&&\left(\mathbf{q}_2[j]\right)_{\vec{a}}^{\vec{b}}=\Bigg(\prod_{i=1}^{j-1} \delta^{b_i}_{a_i}\Bigg) \ 
\frac{d^2}{du^2}W\left[\begin{matrix}
 b_{j-1} & b_j\\
 a_j & a_{j+1}
\end{matrix}\biggl| u\right]\biggr|_{u=0}\ \Bigg(
\prod_{i=j+1}^N \delta^{b_i}_{a_i}\Bigg)-\cr &&-\Bigg( \prod_{i=1}^{j-1}\delta^{b_i}_{a_i} \Bigg)\frac{d}{du}W\left[\begin{matrix}
 b_{j-1} & b_j\\
 a_j & b_{j+1}
\end{matrix}\biggl| u\right]\biggr|_{u=0}\ \frac{d}{du}W\left[\begin{matrix}
 a_{j} & b_{j+1}\\
 a_{j+1} & a_{j+2}
\end{matrix}\biggl| u\right]\biggr|_{u=0}\ \Bigg( \prod_{i=j+2}^N \delta_{a_i}^{b_i} \Bigg).
\end{eqnarray*}

\subsection{Bi-local and $T\bar{T}$ deformations}
In this section we introduce some basic definitions related to both theories: the bi-local deformations of lattice models \cite{LR} and  the $T\bar{T}$ deformations of integrable QFTs \cite {TT}. 

The operator is called bi-local if it is constructed as a sum over products of two local operators, for example, the charges' densities:
\begin{eqnarray}
\mathbf{X}=\sum_{i<j} {\mathbf q_s}[i]{\mathbf q_{r}}[j]\,.
\label{BiLocal}
\end{eqnarray}
Here the indexes $i,j$ denote the cites of the lattice, $r,s$ are labels of the corresponding charges.  

The original definition of the bilocal current-current deformation procedure is given in \cite{LR}, see also \cite{LR1, TTspin, TTspin1}. 
Without going into subtle details, one assumes that  local operators $\mathbf{q}$ and the eigenvectors of the transfer matrix $\ket{\psi}$ are expanding in a series of the deformation parameter $\kappa$
\begin{eqnarray}
&&
\mathbf{{q}}^{(\kappa)}=\sum_m \frac{1}{m!}\mathbf{{q}}_{m}\kappa^m,\cr
&&
\ket{\psi^{(\kappa)}}=\sum_m \frac{1}{m!}\ket{\psi}_m\kappa^m\,.
\label{defcurr}
\end{eqnarray}
Starting from a bare operator (\ref{BiLocal}) and the decomposition (\ref{defcurr}), it is possible to construct deformed operators order-by-order using the equations
\begin{eqnarray*}
&&
\frac{d}{d\kappa}\mathbf{q}^{(\kappa)}=i[\mathbf{X}^{(\kappa)}, \mathbf{q}^{(\kappa)}],\cr
&&
\frac{d}{d\kappa}\ket{\psi^{(\kappa)}}=-i\mathbf{X}^{(\kappa)}\ket{\psi^{(\kappa)}}\,,
\end{eqnarray*}
where the initial conditions are given by the operators and states of an underformed theory ($\kappa=0$). The deformed operator $\mathbf{X}^{(\kappa)}$ should be also understood as a series in $\kappa$
\begin{equation}
    \mathbf{X}^{(\kappa)}=\sum\limits_{i<j}\mathbf{q}^{(\kappa)}_s[i]\mathbf{q}^{(\kappa)}_l[j].
    \label{X}
\end{equation}
It can be proven that the integrability survives under the deformation \cite{LR}. Indeed, if we start from a system which has a set of commuting charges, then the deformed charges will be commuting as well:
\begin{eqnarray*}
&&
[\mathbf{Q}^{(\kappa)}_r, \mathbf{Q}^{(\kappa)}_{s}]=0\,,
\end{eqnarray*}
for arbitrary integers $r,s>0$.

Let us discuss some modification of the prescription for getting new integrable theories with non-local interaction. Recall that the $T\bar{T}$ deformation of a QFT \cite{TT} is defined as the spectral flow of the action (see (\ref{ZS})). An analogous definition can be given in terms of the spectral flow of the Hamiltonian (see for example \cite{DeformedS4}).
\begin{equation}
    \frac{d}{d\kappa}{\mathbf H}^{(\kappa)}=\epsilon^{\mu\nu}\int dx \  {\mathbf J}^{(\kappa)}_{1,\mu}(x){\mathbf J}^{(\kappa)}_{2,\nu}(x),
    \label{mean_derivative}
\end{equation}
where the r.h.s. is constructed from two conserved currents of the deformed theory ${\mathbf J}_{1,\mu}^{(\kappa)}(x)=({\mathbf q}^{(\kappa)}_1(x), {\mathbf j}^{(\kappa)}_1(x))$ and ${\mathbf J}_{2,\mu}^{(\kappa)}(x)=({\mathbf q}^{(\kappa)}_2(x), {\mathbf j}^{(\kappa)}_2(x))$ in the following way
\begin{equation}
     \epsilon^{\mu\nu}{\mathbf J}^{(\kappa)}_{1,\mu}{\mathbf J}^{(\kappa)}_{2,\nu}(x)={\mathbf q}^{(\kappa)}_1(x)\mathbf{j}^{(\kappa)}_2(x)-{\mathbf j}^{(\kappa)}_1(x){\mathbf q}^{(\kappa)}_2(x).
    \label{defo}
\end{equation}
We note that two deformation procedures represented above are almost same but there are some differences. In ref. \cite{LR} the authors considered a 'magnon sector' where all charges act trivially on the corresponding vacuum state. On the other hand, it was shown in ref. \cite{TT} that the energy spectrum for breathers theory (or scaling limit from the regime III), did change as it follows from the Burgers'equation. So for a ferromagnetic regime both approaches give the same results while for the regime III there might be differences. 

\subsection{Bethe ansatz equations}
In order to diagonalize the integrals of motion, we apply the method of the algebraic Bethe ansatz. In general, we follow the procedure introduced in ref. \cite{De_Vega}. 
Eventually, the Bethe ansatz equations resemble those which already appeared
in the work of Bazhanov and Reshetikhin \cite{BR} devoted to studying the unitary RSOS models. However, in our case the parameter $\xi$ (see (\ref{XiParameter})) is not an integer, rather than a rational number. While the string solutions of the BAE strongly depends on the arithmetic properties of this number, or, more exactly, on the correspondent continued fraction.  The explicit form of the Bethe ansatz equations is
\begin{equation}
    \left(\frac{[v_j+\frac{1}{2}]}{[v_j-\frac{1}{2}]}\right)^N=-e^{-2i\omega}\prod\limits_{k=1}^{M}\Bigg(\frac{[v_j-v_k+1]}{[v_j-v_k-1]}\Bigg),
    \label{BAE1}
\end{equation}
for $j=1,..,M$. Here $\omega$ is a real number discussed in \cite{BR}.  Solutions of this system $v_1,..,v_{M}$ are called rapidities. Any set of $M$ rapidities $\vec{v}=(v_1,..,v_{M})$ characterizes a state of the original lattice system $\ket{\psi_{\vec{v}}}$.
The Bethe ansatz equations is a system of nonlinear algebraic equations;  their solutions $v_1,...,v_{M}$ can hardly be found explicitly. Despite this, it is possible to obtain some exact results in the thermodynamic limit when $N,M\rightarrow \infty $.

The system of the Bethe ansatz equations is the necessary condition that the expression
\begin{eqnarray}
&& t(u;\vec{v})= e^{i\omega}\left(\frac{[1-u]}{[1]}\right)^{N}\prod\limits_{k=1}^{M}\frac{[u+v_k+\frac{1}{2}]}{[u+v_k-\frac{1}{2}]} +e^{-i\omega}\left(\frac{[u]}{[1]}\right)^{N}\prod\limits_{k=1}^{M}\frac{[u+v_k-\frac{3}{2}]}{[u+v_k-\frac{1}{2}]}
\label{Eigenvalue}
\end{eqnarray}
is an eigenvalue of the transfer matrix corresponding to a state of the system $\ket{\psi_{\vec{v}}}$
\begin{equation*}
  {\mathbf{T}(u)}\ket{\psi_{\vec{v}}}=t(u;\vec{v})\ket{\psi_{\vec{v}}}.
\end{equation*}
The eigenvalues of all integrals of motion $\mathbf{Q}_r$  can be found using the equation (\ref{Eigenvalue}) according to the definition (\ref{IntQ}). They are represented by  sums over all contribution from quasiparticles which are present in the given state:
\begin{equation}
    \mathbf{Q}_r\ket{\psi_{\vec{v}}}=\sum\limits_{k=1}^{M}q_r(v_k)\ket{\psi_{\vec{v}}}\,.
    \label{q}
\end{equation}
The important for us examples of conserced charges are the Hamiltonian and momentum. We use the following notations for their actions on the Bethe vectors 
\begin{eqnarray*}
&&\mathbf{H}\ket{\psi_{\vec{v}}}=\sum\limits_{k=1}^{M}e(v_k)\ket{\psi_{\vec{v}}}\,,
\cr
&&\mathbf{P}\ket{\psi_{\vec{v}}}=\sum\limits_{k=1}^{M}p(v_k)\ket{\psi_{\vec{v}}}\,.
\end{eqnarray*}
Here the eigenvalues of these operators under the action on the corresponding single-magnon states are given explicitly as
\begin{eqnarray}
&&e(v)=\frac{d}{dv}
\ln{\frac{[v+\frac{1}{2}]}{[v-\frac{1}{2}]}}\,,
\label{e_m}\\
&&p(v)=\frac{1}{i}\ln{\frac{[v+\frac{1}{2}]}{[v-\frac{1}{2}]}}\,.
    \label{p_m}
\end{eqnarray}

\subsection{Deformation of magnon S-matrix}
One of the important for our studying result was obtained in ref. \cite{LR}: the deformation of the integrable lattice model by the bi-local operator (\ref{X}) induces the corresponding change of the asymptotic magnon S-matrix
\begin{eqnarray}
\frac{d}{d\kappa}\ln S^{(\kappa)}(u,v)=i(q_s(v)q_l(u)-q_s(u)q_l(v))\,.
\label{start}
\end{eqnarray}
Note that if one-magnon eigenvalues appearing here do not depend on the deformation parameter $\kappa$, then the differential equation can be easily integrated: 
\begin{eqnarray}
S^{(\kappa)}(u,v)=S(u,v)e^{i\kappa(q_s(v)q_l(u)-q_s(u)q_l(v))}\,.
\label{MagS}
\end{eqnarray}
One sees that the magnon S-matrix is modified by a phase factor of the special form. This is a starting point of our analysis. (Note, however, that we completely ignore subtle problems related with the wrapping phenomena \cite{LR}).

We would like to check if a similar deformation of the S-matrix  would appear in the regime III of the Forrester-Baxter models where the low-energy excitations are elliptic generalizations of breathers. Since the deformation of the S-matrix can be defined in the operator language, it is expected that an analogous differential equation will appear for the breather S-matrix. But in this case the vacuum is non-trivial, therefore, a naive repetition of the equations (\ref{start})-(\ref{MagS}) for breathers will face a problem of  deformation-depending 
eigenvalues. In general, it is impossible to integrate the differential equation (\ref{start}). There should be another reason why the factor analogous to (\ref{MagS}) appears for the breathers' scattering. 

Following the results of ref. \cite{TTspin, TTspin1}, we consider the deformation of our integrable system by the bi-local operator (\ref{X}) constructed from the momentum and energy operators. However, there is a problem with the lattice formulation of this deformation. The exponent of the momentum is the shift operator which shifts hights along the whole lattice row in the horizontal direction. This is why the momentum operator drastically differs from the higher integrals of motion: it can not be written as a sum of local operators. It is not clear then how to apply the scheme of ref. \cite{LR} directly to the lattice model in this case (see, however, the analysis of the deformed Bose gas in ref. \cite{Jiang} where this problem is absent). For this reason, saying that we study a current-current deformation, we mean that we consider the Bethe ansatz equations with the modified magnon S-matrix: 
\begin{eqnarray}
S^{(\kappa)}(u,v)=S(u,v)e^{i\kappa(e(v)p(u)-e(u)p(v))}\,.
\label{magnones}
\end{eqnarray}
Here $e(v)$ and $p(v)$ are one-magnon eigenvalues of the Hamiltonian and momentum. Note that the deformed magnons S-matrix depends on both rapidities $u$ and $v$,  not on their difference as in the case of QFT \cite{TT}. 

\section{String solutions of BAE}
To derive the breather S-matrix, we follow the procedure introduced in \cite{Breather_S_matrix} for the XXZ model which allows to calculate the scattering matrix basing only on the Bethe ansatz equations (\ref{BAE1}) and the string hypothesis. Let us note that it is necessary to modify slightly this procedure in our elliptic situation: the choice of string solutions for the ground and excited states is different. 

The construction is built upon the fact that in the thermodynamic limit the Bethe ansatz equations become linear integral equations. Its explicit solutions allow to find the ground state energy and momentum, the energy and the momentum of particles etc. 

\subsection{String hypothesis}
In studying the states of the system, we follow the string hypothesis \cite{BR, Takahashi_strings,KR}. It is assumed that the solutions of the Bethe ansatz equations in the thermodynamic limit form the complexes (called strings) of special form
\begin{equation}
    u=v-\frac{\xi+1}{2}\nu-\frac{\xi}{2}(n+1-2l),
    \label{strengs}
\end{equation}
where $1<l<n$ is an integer, $n$ is called the length of the string and complex number $v$ is it's center.\footnote{Usually a string is defined as a set of complex numbers with a common real part and equidistantly situated imaginary parts. We will turn to the standard notations when discussing the trigonometric limit.} There are two types of strings which differ in the parity $\nu\in\{0,1\}$: even strings have $\nu=0$ and odd strings have $\nu=1$. We assume that the strings are solutions of the Bethe ansatz equations in the deformed case also since the deformation of the magnon S-matrix does not bring extra poles.

The construction of  states for RSOS models was already discussed in ref. \cite{BR} where the unitary case with integer $\xi>3$ was considered; our situation with rational $\xi=2/(2s+1)$ is non-unitary.

We assume that the ground state consists of of $\frac{N}{2}$ quasiparticles $v_j$, even strings of length $1$. These rapidities are the solutions of the deformed Bethe ansatz equations 
\begin{equation}
    \left(\frac{[v_j+\frac{1}{2}]}{[v_j-\frac{1}{2}]}\right)^L=-e^{-2i\omega}\prod\limits_{k=1}^{\frac{N}{2}}\Bigg(\frac{[v_j-v_k+1]}{[v_j-v_k-1]}f(v_j, v_k)\Bigg),
    \label{BAEgr}
\end{equation}
for $ j=1,..,\frac{N}{2}$. Here, in fact $L=N\delta$, where $\delta$ is the lattice spacing and $L$ is the size of the system. We put $\delta=1$ in our further calculations and restore it at the very end.  Also the deformation of the magnon scattering matrix was introduced according to (\ref{magnones}):
\begin{equation*}
    f(u, v)=e^{i\kappa \left( e(u)p(v)-e(v)p(u) \right)}.
\end{equation*}
Here the deformation parameter $\kappa$ is real, $\kappa=0$ gives undeformed Bethe ansatz equations. The deformation contains quasiparticle's energy $e(u)$ and momentum $p(u)$ which were given in  (\ref{e_m}) and (\ref{p_m}). It is convenient to consider their Fourier series expansions:
\begin{eqnarray}
&&p(u)=-\frac{1}{i}\frac{\xi}{\xi+1}2 \epsilon u-\frac{1}{i}\sum\limits_{m\in Z/\{0 \}}\frac{1}{m}\frac{\sinh{m\epsilon\xi}}{\sinh{m\epsilon(\xi+1)}}e^{2\epsilon um},
    \label{ef}\\
&&
e(u)=-2\epsilon\sum\limits_{m\in Z}\frac{\sinh{m\epsilon\xi}}{\sinh{m\epsilon(\xi+1)}}e^{2\epsilon
     um},
     \label{pf}
\end{eqnarray}
where the parameters $\epsilon$ and $\xi$ were discussed above (see (\ref{XiParameter})). The zeroth term in the energy expansion is understood as a limit $m\to 0$. As for null mode in the momentum operator, it will not affect the computations since in all important steps we use the derivative of the quantities like (\ref{ef}) which is essentially (\ref{pf}).

Let us make a comment on notations. In general, for a function $F(u)$ we define it's Fourier series expansion in the following way 
\begin{equation}
    F(u)=\sum\limits_{m\in Z}F[m]e^{2um\epsilon},
    \label{Fourier_def}
\end{equation}
assuming that $|e^{-2u\epsilon}|<1$. Therefore the corresponding Fourier coefficients are given by integrals
\begin{equation}
    F[m]=\frac{\epsilon}{i\pi}\int\limits_{-\frac{i\pi}{2\epsilon}}^{\frac{i\pi}{2\epsilon}}F(u)e^{-2um\epsilon}du.
\end{equation}
For example, Fourier coefficients (with $m\neq 0$) of guasiparticle's energy and momentum  (\ref{e_m}) and (\ref{p_m})  are 
\begin{eqnarray*}
   && e[m]=-2\epsilon\frac{\sinh{m\epsilon\xi}}{\sinh{m\epsilon(\xi+1)}},\cr &&
    p[m]=-\frac{1}{ m i}\frac{\sinh{m\epsilon\xi}}{\sinh{m\epsilon(\xi+1)}}.
\end{eqnarray*}

Now we return to the system (\ref{BAEgr}). The Bethe ansatz equations can be equivalently written in the logarithmic form:
\begin{equation}
Z_{g}(v_j)=\frac{2\pi i}{L} I_j,    
\label{I_j}
\end{equation}
where integer or half-integer $I_j$ are the so-called Bethe numbers.  Also the useful object, the counting function, was introduced:
\begin{equation}
    Z_{g}(u)=-\ln{\frac{[u+\frac{1}{2}]}{[u-\frac{1}{2}]}}+\frac{1}{L}\sum\limits_{k=1}^{\frac{N}{2}}\ln{\frac{[u-v_k+1]}{[u-v_k-1]}}+i\kappa \Big( e(u)\ P^{(\kappa)}_{g}-  p(u)\ E^{(\kappa)}_{g} \Big), 
    \label{Z}
\end{equation}
where $E^{(\kappa)}_{g}$ and $P^{(\kappa)}_{g}$ denote energy and momentum densities of the ground state. They are given through sums of contributions of all magnons which form the ground state:
\begin{eqnarray}
&&E^{(\kappa)}_{g}=\frac{1}{L}\sum\limits_{k=1}^{\frac{N}{2}}e(v_k), \cr 
&& P^{(\kappa)}_{g}=\frac{1}{L}\sum\limits_{k=1}^{\frac{N}{2}}p(v_k).
\label{def E0 P0}
\end{eqnarray}

We will consider the thermodynamic limit $N\rightarrow \infty$, where a sum over all rapidities turns into an integral with the corresponding root density. For example, the root density of the system (\ref{BAEgr}) is defined using the equation (\ref{I_j}) as following
\begin{equation}
    \rho^{(\kappa)}_{g}(u_j)=\frac{1}{2\epsilon}\lim_{N\rightarrow \infty}\frac{Z_{g}(u_{j+1})-Z_{g}(u_j)}{u_{j+1}-u_j}.
\end{equation}
In the thermodynamic limit the rapidities are assumed to be infinitely close to each other and fill the whole interval $(-\frac{i\pi}{2\epsilon}, \frac{i\pi}{2\epsilon})$ with the continuous root density: 
\begin{equation}
    \rho^{(\kappa)}_{g}(u)=\frac{1}{2\epsilon}\frac{dZ_{g}(u)}{du}.
\end{equation}
Eventually, when going to the thermodynamic limit, one can pass from the sum to an integral according to the following rule
\begin{equation*}
    \frac{1}{L}\sum\limits_{k=1}^{\frac{N}{2}}e(v_k)\rightarrow \frac{\epsilon}{i\pi}\int\limits_{-\frac{i\pi}{2\epsilon}}^{+\frac{i\pi}{2\epsilon}}e(v)\rho^{(\kappa)}(v)dv.
\end{equation*}

\subsection{Ground state energy in the $\kappa=0$ case}
Differentiating the right hand side of ($\ref{Z}$) in the limit $N\rightarrow\infty$, one obtains a linear integral equation for the root density $\rho^{(\kappa)}_{g}(u)$: 
\begin{eqnarray}
    &&\rho^{(\kappa)}_{g}(u)=-\frac{1}{2\epsilon }\frac{d}{du}\ln{\frac{[u+\frac{1}{2}]}{[u-\frac{1}{2}]}}+\frac{1}{2\pi i}\int\limits_{-\frac{i\pi}{2\epsilon}}^{\frac{i\pi}{2\epsilon}}\Bigg(\frac{d}{dv}\ln{\frac{[u-v+1]}{[u-v-1]}}\Bigg)\rho^{(\kappa)}_{g}(v)dv+\cr &&+\frac{i\kappa}{2\epsilon}\Big(\frac{de(u)}{du}\ P^{(\kappa)}_{g}-E^{(\kappa)}_{g}\ \frac{dp(u)}{du}  \Big).
    \label{eq_rho_g}
\end{eqnarray}
Here $E^{(\kappa)}_{g}$ and $P^{(\kappa)}_{g}$ are given in (\ref{def E0 P0}). Fourier transforming their definition, we can formally write
 \begin{equation}
  E^{(\kappa)}_{g}=\sum\limits_{n\in Z}\rho^{(\kappa)}_{g}[-n]\ e[n],
  \label{E0}
\end{equation}
\begin{equation}
    P^{(\kappa)}_{g}=\sum\limits_{n\in Z}\rho^{(\kappa)}_{g}[-n]\ p[n].
    \label{P0}
\end{equation}
Then the root density of the ground state can be found by successive approximations. We provide this explicit calculation below. In the undeformed Forrester-Baxter case the Fourier coefficient of the ground state density is 
\begin{equation}
    \rho^{(0)}_{g}[m]=\frac{1}{2\cosh{m\epsilon}}\,.
    \label{rho0g}
\end{equation}
Then one immediately obtains the ground state energy for the Forrester-Baxter model without deformation 
\begin{equation}
   \mathcal{H}=-2\epsilon\sum\limits_{m\in Z}\frac{\sinh{m\epsilon\xi}}{\sinh{m\epsilon(\xi+1)}}\frac{1}{2\cosh{m\epsilon}}\,.
   \label{Egr}
\end{equation}
We can compare this expression with the partition function per site of the eight vertex model \cite{Baxter:book,Baxter1972} for the corresponding crossing parameter (see eq. 4.6 in \cite{LHP} in our notations).
Also one can think of (\ref{Egr}) as an analytic continuation from integer to fractional $\xi$ of the expression of the ground state energy, which was given in \cite{BR}. But the choice of strings for the ground state drastically differs from their choice in unitary regime. 

In the deformed situation the answers given here become $\kappa$-dependent. We consider this case in details in the next subsections.

\subsection{Solving the deformed BAE}
In this section we solve the linear integral equation for the deformed ground state density (\ref{eq_rho_g}). Recall the equation
\begin{eqnarray*}
    &&\rho^{(\kappa)}_{g}(u)=-\frac{1}{2\epsilon }\frac{d}{du}\ln{\frac{[u+\frac{1}{2}]}{[u-\frac{1}{2}]}}+\frac{1}{2\pi i}\int\limits_{-\frac{i\pi}{2\epsilon}}^{\frac{i\pi}{2\epsilon}}\Bigg(\frac{d}{dv}\ln{\frac{[u-v+1]}{[u-v-1]}}\Bigg)\rho^{(\kappa)}_{g}(v)dv+\cr &&+\frac{i\kappa}{2\epsilon}\Big(\frac{de(u)}{du}\ P^{(\kappa)}_{g}-E^{(\kappa)}_{g}\ \frac{dp(u)}{du}  \Big).
\end{eqnarray*}
As usual, a linear integral equation of this kind can be solved by the Fourier series expansion. In terms of the Fourier coefficients ($m\neq 0$) this equation has the following form
\begin{eqnarray*}
&&\rho^{(\kappa)}_g[m]=\frac{1}{2\cosh{m\epsilon}}-\frac{\kappa}{2\cosh{m\epsilon}} \Big( 2m\epsilon \cdot i\sum\limits_{n\in Z}\rho^{(\kappa)}_g[-n]\ p[n]-\sum\limits_{n\in Z}\rho^{(\kappa)}_g[-n]\ e[n] \Big),
\end{eqnarray*}
where we used $E^{(\kappa)}_{g}$ and $P^{(\kappa)}_g$ given in (\ref{E0}) and (\ref{P0}). We look for the solution of this equation in the form of a power series in $\kappa$ 
\begin{equation}
   \rho^{(\kappa)}_g[m]=\rho_{g,0}[m]+\kappa \rho_{g,1}[m]+\kappa^2\rho_{g,2}[m]+... 
   \label{PertFB}
\end{equation}
Here the  zeroth order corresponds to the ground state density in the pure case (\ref{rho0g}). Then for the first order the equation is given as following
\begin{equation*}
  \rho_{g,1}[m]=-\frac{1}{2\cosh{m\epsilon}}\Big( 2m\epsilon  \cdot i\sum\limits_{n\in Z}\rho_{g,0}[-n]\ p[n] -\sum\limits_{n\in Z}\rho_{g,0}[-n]\ e[n]\Big) \,.
\end{equation*}
Note that the first term in the right hand side vanishes. The second term in the brackets is exactly the energy $\mathcal H$ of the undeformed ground state  (\ref{Egr}). Then in the first order in $\kappa$ we get a very simple result
\begin{equation*}
  \rho_{g,1}[m]=\frac{\mathcal{H}}{2\cosh{m\epsilon}}\,.
\end{equation*}
All other terms in the expansion (\ref{PertFB}) can be found exactly by this procedure: to obtain the equation for $\rho_{g,j}[m]$ one should put $\rho_{g, j-1}[m]$ in the definition of $E^{(\kappa)}_g$ and $P^{(\kappa)}_g$. Then one finds 
\begin{equation*}
    \rho_{g,j}[m]=\frac{1}{2\cosh{m\epsilon}}\mathcal{H}^j\,.
\end{equation*}
Collecting all terms together, one finds that the ground state density changed in a very neat way:
\begin{equation*}
    \rho_g^{(\kappa)}[m]=\frac{1}{2\cosh{m\epsilon}}\sum_{j=0}^{\infty}\mathcal{H}^j\kappa^j\,.
\end{equation*}
This is the result of the special form of the deformation: it is constructed from integrals of motion which in turn are logarithmic derivatives of the transfer matrix (\ref{IntQ}).

If the parameter $\kappa$ is small, $|\kappa\mathcal{H}|<1$, the ground state density multiplies by the factor which depends on the energy density of the pure ground state:  
\begin{equation}
    \rho^{(\kappa)}_{g}[m]=\frac{1}{1-\kappa\mathcal{H}}\ \frac{1}{2\cosh{m\epsilon}}.
    \label{re4}
\end{equation}
Note that all this machinery works effectively for finding root density of the excited state described below.

Now it is possible to find $E^{(\kappa)}_{g}$ and $P^{(\kappa)}_{g}$ explicitly. Putting (\ref{re4}) in (\ref{E0}), one finds that the density of the ground state energy in the deformed case is 
\begin{equation}
    E^{(\kappa)}_g=\frac{\mathcal{H}}{1-\kappa \mathcal{H}}\,.
    \label{change_e_g}
\end{equation}
Also from (\ref{P0}) follows that the total momentum of the ground state is zero:
\begin{equation*}
    P_g^{(\kappa)}=0.
\end{equation*}
We point out that the deformation of the Bethe ansatz equations (\ref{BAEgr}) led to the same shift of the energy density which  was observed in $T\bar{T}$-deformed QFTs \cite{TT}. 
In connection with this observation, let us discuss the obtained result. The change of the ground state density effectively means that the size of the system changed:
\begin{equation*}
   L^{(\kappa)}=L(1-\kappa\mathcal{H}).
\end{equation*}
As the number of sites of the lattice is fixed, this change is equivalent to the change of the lattice spacing $\delta$:
\begin{equation}
    \delta^{(\kappa)}=\delta(1-\kappa\mathcal{H}).
\end{equation}
From the last observation follows that one can expect the change of masses of particles which indeed will be demonstrated considering the scaling limit. 

\subsection{Excited states}
It is known that in an integrable case in $1+1$ dimensions many-particle collisions are reduced to a set of consecutive pairwise collisions. Therefore, many-particle S-matrix elements have special structure: they factorize into the product of two-particle S-matrix elements. It means that it is enough to consider only two-particle scattering processes. Taking this into account, we consider an excited which consists of two odd strings $u_1, u_2$ 
\begin{eqnarray}
    && u_1=w_1-\frac{\xi+1}{2}-\frac{\xi}{2}(n_1+1-2l_1), \cr && u_2=w_2-\frac{\xi+1}{2}-\frac{\xi}{2}(n_2+1-2l_2), 
    \label{br1}
\end{eqnarray}
above the Dirac sea formed by even strings $v_j$ with lengths equal to 1. For such a state  the system of the Bethe ansatz equations have the form: 
\begin{eqnarray}
     &&\left(\frac{[v_j+\frac{1}{2}]}{[v_j-\frac{1}{2}]}\right)^L=-e^{-2i\omega}\prod\limits_{k=1}^{M}\frac{[v_j-v_k+1]}{[v_j-v_k-1]}f(v_j,v_k)\cdot \cr &&\cdot \prod\limits_{l_1=1}^{n_1}\frac{[v_j-u_1+1]}{[v_j-u_1-1]}f(v_j,u_1) \prod\limits_{l_2=1}^{n_2}\frac{[v_j-u_2+1]}{[v_j-u_2-1]}f(v_j,u_2),
     \label{newS4}
\end{eqnarray}
where $M\rightarrow\infty$ is the number of magnons in the Dirac sea. 

It is useful to write the system (\ref{newS4}) in the equivalent logarithmic form
\begin{equation*}
    Z(v_j)=\frac{2\pi i}{L} I_j,
\end{equation*}
where $I_j$ are the Bethe numbers which differ from the previous case (\ref{I_j}) but we use the same letter for convenience. The corresponding counting function is 
\begin{eqnarray*}
&&Z(u)=-\ln{\frac{[u+\frac{1}{2}]}{[u-\frac{1}{2}]}}+\frac{1}{L}\sum\limits_{k=1}^{M}\ln{\frac{[u-v_k+1]}{[u-v_k-1]}} +i\kappa \Big(e(u)\ P^{(\kappa)}- p(u)\ E^{(\kappa)} \Big)+\cr && +\frac{1}{L}\sum\limits_{p=1}^2\sum\limits_{l_p=1}^{n_p}\ln{\frac{[u-w_p+\frac{1-\xi}{2}+\frac{\xi}{2}(n_p+1-2l_p)]}{[u-w_p-\frac{1-\xi}{2}+\frac{\xi}{2}(n_p+1-2l_p)]}}+\cr &&+\frac{i\kappa}{L}\sum\limits_{p=1}^2\sum\limits_{l_p=1}^{n_p}\Bigg(e(u)p\Big(w_p-\frac{\xi+1}{2}-\frac{\xi}{2}(n_p+1-2l_p)\Big) -\cr &&- e\Big(w_p-\frac{\xi+1}{2}-\frac{\xi}{2}(n_p+1-2l_p)\Big) p(u) \Bigg).
\end{eqnarray*}
Here we used the periodicity property $[u+\xi+1]=-[u]$ and the minus simply led to the redefinition of the Bethe numbers when taking logarithms. The total energy and momentum densities of the excited state are defined in the following way
\begin{eqnarray}
   && E^{(\kappa)}=\sum\limits_{n\in Z}\rho^{(\kappa)}[-n]\ e[n],\cr &&
   P^{(\kappa)}=\sum\limits_{n\in Z}\rho^{(\kappa)}[-n]\ p[n],
   \label{EP}
\end{eqnarray}
where the density of roots of the system (\ref{newS4}) describes the Dirac sea, distorted by the presence of the strings (\ref{br1}). It is defined through the counting function
\begin{equation}
    \rho^{(\kappa)}(u)=\frac{1}{2\epsilon}\frac{dZ(u)}{du}.
    \label{dens}
\end{equation}
 Differentiating the right hand side of (\ref{newS4}) and passing from the sum to an integral, we obtain a linear integral equation for this density:
\begin{eqnarray}
    && \rho^{(\kappa)}(u)=-\frac{1}{2\epsilon}\frac{d}{du}\ln{\frac{[u+\frac{1}{2}]}{[u-\frac{1}{2}]}}+\frac{1}{2\pi i}\int\limits_{-\frac{i\pi}{2\epsilon}}^{+\frac{i\pi}{2\epsilon}}\Bigg(\frac{d}{dv}\ln{\frac{[u-v+1]}{[u-v-1]}}  \Bigg)\rho^{(\kappa)}(v)dv+\cr&& + \frac{i\kappa}{2\epsilon}\Big(\frac{de(u)}{du} \ P^{(\kappa)} -\frac{dp(u)}{du} \ E^{(\kappa)}  \Big)+\frac{1}{2\epsilon L}\sum\limits_{p=1}^2\sum\limits_{l_p=1}^{n_p}\frac{d}{du}\ln{\frac{[u-w_p+\frac{1-\xi}{2}+\frac{\xi}{2}(n_p+1-2l_p)]}{[u-w_p-\frac{1-\xi}{2}+\frac{\xi}{2}(n_p+1-2l_p)]}}+\cr &&+\frac{i\kappa}{2\epsilon L}\sum\limits_{p=1}^2\sum\limits_{l_p=1}^{n_p}\Big(\frac{de(u)}{du}\cdot p(w_p-\frac{\xi+1}{2}-\frac{\xi}{2}(n_p+1-2l_p))-\cr &&-e(w_p-\frac{\xi+1}{2}-\frac{\xi}{2}(n_p+1-2l_p))\cdot \frac{dp(u)}{du}  \Big).
 \end{eqnarray}
Similarly to the equation for the ground state density (\ref{eq_rho_g}), the solution of this equation can be found in the form of a power series in $\kappa$. The zeroth order gives the pure Forrester-Baxter answer, using which one can fix the first order, etc. Finally one finds the Fourier coefficient of the deformed density (\ref{dens})
\begin{eqnarray}
   && \rho^{(\kappa)}[m]=\frac{1}{1-\kappa\mathcal{H}}\ \frac{1}{2\cosh{m\epsilon}}-\frac{1}{L}\frac{\cosh{m\epsilon\xi}}{\cosh{m\epsilon}}\sum\limits_{p=1}^2\frac{\sinh{m\epsilon\xi n_p}}{\sinh{m\epsilon\xi}}e^{-2m\epsilon w_p}-\cr && -\frac{\kappa}{L}\frac{1}{1-\kappa\mathcal{H}}\ \frac{1}{2\cosh{m\epsilon}}\Big(2m\epsilon\cdot i\sum\limits_{p=1}^2\mathcal{P}_{n_p}(w_p)-\sum\limits_{p=1}^2\mathcal{E}_{n_p}(w_p)  \Big).
   \label{density}
\end{eqnarray}
Here we introduced new functions  $\mathcal{E}_{n}(u)$ and $\mathcal{P}_n(u)$   determined by their Fourier coefficients
\begin{eqnarray*}
 &&\mathcal{E}_{n}[m]=-2\epsilon\frac{\cosh{m\epsilon(n\xi+1)}}{\cosh{m\epsilon}},\cr &&
   \mathcal{P}_{n}[m]=-\frac{1}{i m}\frac{\cosh{m\epsilon(n\xi+1)}}{\cosh{m\epsilon}}.
\end{eqnarray*}
As we will see, these functions are the undeformed energy and momentum of a particle.
Note that $\rho[m]$ also depends on parameters $w_1$ and $w_2$ which are the centers of the strings which appear in the excited state. We will omit the explicit dependence on these parameters in our notations. 

Putting this root density in (\ref{EP}), one can obtain the energy and momentum densities of the excited state:
\begin{eqnarray}
   && E^{(\kappa)}+\frac{1}{L}\sum\limits_{p=1}^2e_{n_p}(w_p)=\frac{\mathcal{H}}{1-\kappa\mathcal{H}}+\frac{1}{L}\frac{1}{1-\kappa\mathcal{H}}\sum\limits_{p=1}^{2}\mathcal{E}_{n_p}(w_p),
   \cr && P^{(\kappa)}+\frac{1}{L}\sum\limits_{p=1}^2p_{n_p}(w_p)=\frac{1}{L}\frac{1}{1-\kappa\mathcal{H}}\sum\limits_{p=1}^2\mathcal{P}_{n_p}(w_p),
\label{Ha_distor}
\end{eqnarray}
where $e_n(u)$ and $p_n(u)$ are the energy and momentum of a string of length $n$: 
\begin{eqnarray*}
   && e_{n}(u)=\frac{d}{du}\ln{\frac{[u-\frac{n\xi}{2}]}{[u+\frac{n\xi}{2}]}},
   \cr && p_n(u)=\frac{1}{i}\ln{\frac{[u-\frac{n\xi}{2}]}{[u+\frac{n\xi}{2}]}}.
\end{eqnarray*}

These results indicate that the energy and momentum of the excited state are composed from the energy and momentum of the Dirac sea and the contribution from two breathers. The Fourier coefficients of the deformed breather energy and momentum are
\begin{eqnarray}
     &&\mathcal{E}^{(\kappa)}_{n}[m]=-\frac{2\epsilon}{1-\kappa\mathcal{H}}\ \frac{\cosh{m\epsilon(n\xi+1)}}{\cosh{m\epsilon}},
      \cr && \mathcal{P}^{(\kappa)}_n[m]=-\frac{1}{i m}\frac{1}{1-\kappa\mathcal{H}}\ \frac{\cosh{m\epsilon(n\xi+1)}}{\cosh{m\epsilon}}.
     \label{br_energy}
\end{eqnarray}
where the number $n$ is the length of the corresponding string excitation.

Now it is easy to see that breather energy and momentum are related by an important property
\begin{equation}
    \mathcal{E}^{(\kappa)}_n(u)=i\frac{d}{du}\mathcal{P}^{(\kappa)}_n(u).
    \label{BrMom}
\end{equation}
The explicit expressions of breather energy and momentum in the undeformed case can be computed using  (\ref{br_energy}), where one should put $\kappa=0$:
\begin{equation}
    \mathcal{P}_a(u)=-\frac{1}{i}\ln{\frac{\Theta_2\Big( \frac{\pi}{2}(u+\frac{a\xi}{2}) |\frac{i\pi}{2\epsilon}\Big)\Theta_1\Big( \frac{\pi}{2}(u-\frac{a\xi}{2}) |\frac{i\pi}{2\epsilon}\Big)}{\Theta_1\Big(\frac{\pi}{2}(u+\frac{a\xi}{2})  |\frac{i\pi}{2\epsilon}\Big)\Theta_2\Big(\frac{\pi}{2}(u-\frac{a\xi}{2})|\frac{i\pi}{2\epsilon}  \Big)}},
\label{mathcalP}
\end{equation}
\begin{equation}
    \mathcal{E}_a(u)=-\frac{ d } { du }\ln{\frac{\Theta_2\Big( \frac{\pi}{2}(u+\frac{a\xi}{2})|\frac{i\pi}{2\epsilon} \Big)\Theta_1\Big( \frac{\pi}{2}(u-\frac{a\xi}{2})|\frac{i\pi}{2\epsilon} \Big)}{\Theta_1\Big(\frac{\pi}{2}(u+\frac{a\xi}{2}) |\frac{i\pi}{2\epsilon} \Big)\Theta_2\Big(\frac{\pi}{2}(u-\frac{a\xi}{2})  |\frac{i\pi}{2\epsilon}\Big)}}.
\label{mathcalE}
\end{equation}
The deformation with $\kappa\neq 0$ simply lead to the multiplication by the factor which we have already seen
\begin{eqnarray}
&&
\mathcal{E}_a^{(\kappa)}(u)=\frac{\mathcal{E}_a(u)}{1-\kappa \mathcal{H}}\,,\cr
&&
\mathcal{P}_a^{(\kappa)}(u)=\frac{\mathcal{P}_a(u)}{1-\kappa \mathcal{H}}\,.
\label{PE}
\end{eqnarray}
The next step is to exploit the Bethe ansatz equations, which describe the breathers. We write the equation for the first string $u_1$ from (\ref{br1}) and form the product $\prod\limits_{l_1=1}^{n_1}$ of both parts of it since the equation is satisfied for the string as a whole:
\begin{eqnarray}
&&\prod\limits_{l_1=1}^{n_1}\left(\frac{[w_1-\frac{\xi}{2}-\frac{\xi}{2}(n_1+1-2l_1)]}{[w_1+\frac{\xi}{2}-\frac{\xi}{2}(n_1+1-2l_1)]}\right)^L=\cr && =-e^{-2i\theta}\prod\limits_{l_1=1}^{n_1}\Bigg(\prod\limits_{k=1}^{M}\frac{[w_1-\frac{\xi}{2}(n_1+1-2l_1)-v_k+\frac{1-\xi}{2} ]}{[w_1-\frac{\xi}{2}(n_1+1-2l_1)-v_k-\frac{1-\xi}{2}]}f(w_1-\frac{\xi+1}{2}-\frac{\xi}{2}(n_1+1-2l_1),v_k)\cdot \cr && \cdot
\prod\limits_{l_2=1}^{n_2}\frac{[w_1-w_2-\frac{\xi}{2}(n_2-n_1-2l_2+2l_1)+1]}{[w_1-w_2+\frac{\xi}{2}(n_2-n_1-2l_2+2l_1)-1]}\cdot \cr &&\cdot f(w_1-\frac{\xi+1}{2}-\frac{\xi}{2}(n_1+1-2l_1),w_2-\frac{\xi+1}{2}-\frac{\xi}{2}(n_2+1-2l_2))\Bigg).
\label{fB}
\end{eqnarray}
Here $j=1,..,\frac{N}{2}$ and we used the property that $[u+\xi+1]=-[u]$. After taking logarithms, these equations can be written in the equivalent form
\begin{equation*}
    \tilde Z(v_j)=\frac{2\pi i}{L} I_j.
\end{equation*}
 The Bethe numbers $I_j$ are related to the system (\ref{fB}) and $\tilde Z(u)$ is the corresponding counting function:
\begin{eqnarray*}
    &&\tilde Z(u)= -\ln{\frac{[u-\frac{n_1\xi}{2}]}{[u+\frac{n_1\xi}{2}]}} +\sum\limits_{l_1=1}^{n_1}\Bigg[\frac{1}{L}\sum\limits_{k=1}^{M}\ln{\frac{[u-\frac{\xi}{2}(n_1+1-2l_1)-v_k+\frac{1-\xi}{2}]}{[u-\frac{\xi}{2}(n_1+1-2l_1)-v_k-\frac{1-\xi}{2}]}} +\cr && 
    +i\kappa \Bigg(e\big(u-\frac{\xi}{2}(n_1+1-2l_1)-\frac{\xi+1}{2}\big)\cdot P^{(\kappa)}- E^{(\kappa)}\cdot p\big(u-\frac{\xi}{2}(n_1+1-2l_1)-\frac{\xi+1}{2}\big) \Bigg)+\cr && +\frac{1}{L}
    \sum\limits_{l_2=1}^{n_2}\ln{\frac{[u-\frac{\xi}{2}(n_1+1-2l_1) -w_2+\frac{\xi}{2}(n_2+1-2l_2) +1]}{[u-\frac{\xi}{2}(n_1+1-2l_1) -w_2+\frac{\xi}{2}(n_2+1-2l_2) -1]}} +\cr && +  \frac{i\kappa}{L}\Bigg(e\big(u-\frac{\xi}{2}(n_1+1-2l_1)-\frac{\xi+1}{2}\big)\cdot \sum\limits_{l_2=1}^{n_2} p(w_2-\frac{\xi}{2}(n_2+1-2l_2)-\frac{\xi+1}{2})-\cr && -\sum\limits_{l_2=1}^{n_2} e(w_2-\frac{\xi}{2}(n_2+1-2l_2)-\frac{\xi+1}{2})\cdot p\big(u-\frac{\xi}{2}(n_1+1-2l_1)-\frac{\xi+1}{2}\big) \Bigg) \Bigg].
\end{eqnarray*}

The density of the roots of the system (\ref{fB}) is defined as usual through the counting function
\begin{equation}
    \sigma^{(\kappa)}(u)=\frac{1}{2\epsilon}\frac{d\tilde Z(u)}{du}.
    \label{sZ}
\end{equation}
After differentiating the right hand side of $\tilde Z(u)$,  one should pass from the sum $\sum\limits_{k=1}^M$ to an integral where one should use the root density of the excited state (\ref{density}). After lengthy but direct calculations the Fourier coefficient of $\sigma^{(\kappa)}(u)$ can be represented in a nice form 
\begin{eqnarray}
&&\sigma^{(\kappa)}[m]=\mathcal{E}_n^{(\kappa)}[m]+\frac{1}{L}\frac{\sinh{m\epsilon\xi n_1}}{\sinh{m\epsilon\xi}}\frac{\sinh{m\epsilon\xi n_2}}{\sinh{m\epsilon\xi}}\frac{\cosh{m\epsilon(2\xi-1)}}{\cosh{m\epsilon}}e^{-2m\epsilon w_2}+\cr &&+\frac{1}{L}\Bigg(\frac{\sinh{m\epsilon\xi n_1}}{\sinh{m\epsilon\xi}}  \Bigg)^2\frac{\sinh{2m\epsilon\xi}}{\sinh{m\epsilon(\xi+1)}}\frac{\cosh{m\epsilon(\xi+1)}}{\cosh{m\epsilon}}e^{-2m\epsilon w_1}+\cr &&+\frac{\kappa}{L}\frac{ \mathcal{E}_{n_1}[m]}{1-\kappa\mathcal{H}}\Bigg( 2m\epsilon\cdot i\sum\limits_{p=1}^2\mathcal{P}_{n_p}(w_p)- \sum\limits_{p=1}^2\mathcal{E}_{n_p}(w_p)  \Bigg)
\label{bar_sigma}
\end{eqnarray}
This expression is a basic one in the derivation of the scattering matrix both in the deformed and pure cases.

\section{S-matrix}
In this section we provide exact computation of the scattering matrix of a Forrester-Baxter model $(2,2s+1)$.
Such a theory has only one vacuum so there are no kinks which interpolate between different vacua and have, in general, non-diagonal scattering. They disappeared from the spectrum due to the quantum group reduction.  It is expected that the deformed S-matrix is also diagonal. 
In this simplified situation we suppose that the string excitation of length $n$ corresponds to a deformed elliptic higher breather of number $n$ with energy $\mathcal{E}^{(\kappa)}_n(u)$ and momentum $\mathcal{P}^{(\kappa)}_n(u)$ as given in  (\ref{PE}).


We expect that the deformed S-matrix has an additional CDD-like factor. Our main goal is to check the exact form of this factor and trace out how the additional multiple in the magnon S-matrix transforms into the additional CDD factor \cite{CDD, TT} for scattering of physical  particles.

\subsection{Breather-like scattering in the elliptic case}
We denote $\mathcal{S}^{(\kappa)}_{n_1, n_2}$ the phase which describes the scattering of a breather of number $n_1$ and a breather of number $n_2$.
Note that it is necessary to take into account that the physical vacuum in the regime III in Forrester-Baxter model is non-trivial. The vacuum is the Dirac sea of quasi-particles which also contribute to the scattering amplitude \cite{Andrei,Korepin,Hrachik} and the main problem is to take into account this contribution. 
We define the breather-like scattering matrix as the momentum quantization condition \cite{Andrei}
\begin{equation}
    \Big(e^{i\mathcal{P}^{(\kappa)}_{n_1}(w_1)L}\mathcal{S}^{(\kappa)}_{n_1, n_2}(w_1, w_2)-1\Big)\ket{w_1, w_2}=0,
\end{equation}
where $\ket{w_1,w_2}$ is a state of two particles with rapidities $w_1$ and $w_2$; the breather momentum $\mathcal{P}^{(\kappa)}_n(u)$ is given in (\ref{PE}).

It is useful to consider the identity 
\begin{equation*}
i\frac{d}{du}\mathcal{P}^{(\kappa)}_{n_1}(u)-\mathcal{E}^{(\kappa)}_{n_1}(u)+\sigma^{(\kappa)}(u)-\frac{d \tilde Z(u)}{du}=0,
\end{equation*}
which immediately follows from equations (\ref{BrMom}) and (\ref{sZ}). Integrating this identity with respect to $u$ from $-\infty$ to $w_1$, we obtain
\begin{equation}
   \mathcal{ S}^{(\kappa)}_{n_1,n_2}(w_1, w_2)=C \ \exp\Bigg[ \int\limits_{-\infty}^{w_1}\Big(\sigma^{(\kappa)}(u)-\mathcal{E}^{(\kappa)}_{n_1}(u) \Big) du\Bigg],
    \label{S_def}
\end{equation}
where $C$ is a constant: we exclude any factors which do not depend on $w_1$ and $w_2$ from the S-matrix. The Fourier coefficient of the integrand $\sigma^{(\kappa)}[m]-\mathcal{E}^{(\kappa)}_{n_1}[m]$ can be constructed using (\ref{br_energy}) and (\ref{bar_sigma}).  Then the nontrivial contribution to the S-matrix is
\begin{eqnarray}
&&\ln\mathcal{S}^{(\kappa)}_{n_1,n_2}(w_1,w_2)= \frac{1}{2L\epsilon}\sum\limits_{m\in Z/\{0\}}\frac{\cosh{m\epsilon(2\xi-1)}}{\cosh{m\epsilon}}\frac{\sinh{m\epsilon\xi n_1}}{\sinh{m\epsilon\xi }}\frac{\sinh{m\epsilon\xi n_2}}{\sinh{m\epsilon\xi}}\frac{e^{2m\epsilon (w_1-w_2)}}{m}+\cr && +\frac{\kappa}{2L\epsilon}\frac{1}{1-\kappa \mathcal{H}}\sum\limits_{m\in Z/ \{0\}}\mathcal{E}_{n_1}[m]\Bigg( 2\epsilon i \sum\limits_{p=1}^2\mathcal{P}_{n_p}(w_p)-\frac{1}{m}\sum\limits_{p=1}^2\mathcal{E}_{n_p}(w_p) \Bigg)e^{-2w_1m\epsilon}.
\label{Sanswer}
\end{eqnarray}
This is an exact answer for the deformed scattering matrix. Let us rewrite it in a standard form. First, consider the pure Forrester-Baxter case with $\kappa=0$. The answer for the scattering of the elliptic analogues of breathers comes from the first line of equation (\ref{Sanswer})
\begin{equation}
\mathcal{S}_{a, b}(w_1, w_2)= \prod\limits_{l_1=1}^{a}\prod\limits_{l_2=1}^{b}\mathcal{S}_{1,1}\Big(w_1-w_2-\frac{\xi}{2}(n_1-2l_1-n_2+2l_2)\Big). 
\label{SHigherBreathers1}
\end{equation}
It is clear from this expression that scattering has a factorized form which is expected for the S-matrix of higher breathers. We see that general matrix is built from the two-particle one - the scattering matrix for basic particles related to strings of lengths $a=b=1$:
\begin{equation}
\mathcal{S}_{1,1}(w)=\frac{\Theta_1\left(\pi\frac{w+\xi}{2}|\frac{i\pi}{2\epsilon}\right)\Theta_2\left(\pi\frac{w-\xi}{2}|\frac{i\pi}{2\epsilon}\right)}{\Theta_1\left(\pi\frac{w-\xi}{2}|\frac{i\pi}{2\epsilon}\right)\Theta_2\left(\pi\frac{w+\xi}{2}|\frac{i\pi}{2\epsilon}\right)}.
\label{SFirstBreather}\end{equation}
Then the general answer for the deformed S-matrix of excitations corresponding to the string solutions of the BAE can be writen as following:
\begin{equation}
    \mathcal{S}^{(\kappa)}_{n,m}(w_1,w_2)=\mathcal{S}_{n,m}(w_1,w_2)e^{ \frac{i\kappa}{1-\kappa \mathcal{H}}\big(\mathcal{P}_{n}(w_1)\mathcal{E}_{m}(w_2)-\mathcal{E}_{n}(w_1)\mathcal{P}_{m}(w_2)  \big)}.
    \label{finish}
\end{equation}
Note that the extra multiple appearing in the S-matrix (\ref{finish}) is different from the magnons' case (\ref{magnones}) since it is constructed from the undeformed breather energy $\mathcal{E}_n(u)$ and momentum $\mathcal{P}_n(u)$, not from magnon ones. Also there is an additional multiple $1/(1-\kappa \mathcal{H})$.  

Since the dispersion relation for elliptic breathers is not relativistic, this extra multiple is not the usual relativistic CDD factor: it depends on both rapidities of scattering particles,  not on their difference. However, it has all other basic properties: it does not destroy the Yang-Baxter as well as the bootstrap equations and does not have extra poles in the physical region. In the next sections we will show that the genuine CDD multiple appears in the scaling limit.

\subsection{Scaling limit vs  $T\bar{T}$ perturbed QFT}
In this section we consider the scaling limit of the deformed Forrester-Baxter model and compare our main results with those that are known for $T\bar{T}$-deformed QFTs from ref. \cite{TT}.

In the scaling limit the Forrester-Baxter model in the regime III is described by the Restricted sine-Gordon model with  $\xi=2/(2s-1)$\cite{AlZam,LeClair,SmirnovRG}.
The theory can be defined as a $\Phi_{13}$ perturbation of the minimal models of BPZ. Alternatively, it can be described as a scattering theory of $s-1$ neutral breathers with a specific mass ratio and exact S-matrix \cite{Cardy,AlZam,Freund}. Let us recall that the conformal limit is obtained by sending $\epsilon\to 0$. In this limit the lattice model becomes the minimal CFT. In a vicinity of the critical point in the region of massive QFT the model develops a finite correlation length which is related with the inverse mass. 

Going to the limit $\epsilon\to 0$ in (\ref{mathcalP})-(\ref{mathcalE}) near the energy minimum and performing the change $\pi u=-i\beta$, one finds the undeformed breatherenergy and momentum:
\begin{eqnarray}
   &&\mathcal{E}_n(\beta)=8\frac{e^{-{\pi^2}/{2\epsilon}}}{\delta}\sin{\frac{\pi n \xi}{2}}\cosh{\beta} \,,\cr &&
   \mathcal{P}_n(\beta)=8\frac{e^{-{\pi^2}/{2\epsilon}}}{\delta}\sin{\frac{\pi n\xi}{2}}\sinh{\beta}\,.
   \label{etr}
\end{eqnarray}
We reconstructed here the lattice step $\delta$ which has a dimension of the length. The scaling limit is obtained as well by sending $\delta\to 0$ keeping masses finite. Also we rescaled the Hamiltonian $\mathcal{H}\rightarrow\frac{1}{\pi}\mathcal{H}$, therefore the breather energy and momentum are related as following: 
\begin{equation*}
    \mathcal{E}_n(\beta)=\frac{d}{d\beta}\mathcal{P}_n(\beta).
\end{equation*}
The mass of the first breather, which corresponds to a string excitation of  length $n=1$, follows then from (\ref{etr}):
\begin{equation}
    m_1=8\frac{e^{-{\pi^2}/{2\epsilon}}}{\delta}\sin{\frac{\pi  \xi}{2}}\,. 
\end{equation}
As was mentioned above, the quantity $m_1$  is finite. A higher breather, which corresponds to a string of length $n$, has the mass
\begin{equation}
    m_n=m_1\frac{\sin{\frac{\pi n\xi}{2}}}{\sin{\frac{\pi\xi}{2}}}, \quad 1\leq n\leq s-1\,.
\end{equation}
Now let us discuss the scaling limit of the ground state energy density (\ref{Egr}). The exact answer for the Restricted sine-Gordon model is known from the TBA approach \cite{AlZam,AlZamTBA}. It also can be computed using the method of ref. \cite{DeVega} or obtained from  the partition function of the eight vertex model \cite{Baxter1972} as was commented in ref. \cite{LZ}. To derive this expression from our answer we provide a following procedure. First, we perform the Poisson resummation and obtain the expansion
\begin{equation}
    \frac{1}{\pi}\mathcal{H}=-\frac{2}{\pi \delta^2}\sum\limits_{m\in Z}\int\limits_{-\infty}^{+\infty}e^{i\frac{2\pi mt}{\epsilon}}\frac{\sinh{t\xi}}{\sinh{t(\xi+1)}}\frac{1}{2\cosh{t}}dt.
\label{EllSer}
\end{equation}
The leading contribution to this sum for small $\epsilon$ is given by the term $m=0$. It is a well-known expression for the energy of the ground state in the critical RSOS chain:
\begin{equation}
    \frac{1}{\pi}\mathcal{H}_0=-\frac{1}{\pi \delta^2}\int_{-\infty}^{\infty}dt\frac{\sinh{t\xi}}{\sinh{t(\xi+1)}}\frac{1}{\cosh{t}}\,.
    \label{H=H_0+}
\end{equation}
This quantity become infinitely large in the continuous limit since the lattice step $\delta$ is going to zero. For this reason there is a standard prescription to define the ground state energy to be zero in the critical point. Therefore, going from the lattice to a continuous model, we have to redefine the Hamiltonian by subtracting the constant (\ref{H=H_0+}):
\begin{equation*}
   \mathbf{H}\rightarrow \mathbf{H}-N\mathcal{H}_0.
\end{equation*}
Then the essential contribution to the ground energy in the scaling theory comes from the $m=\pm 1$  terms of the sum (\ref{EllSer}).  To calculate the $m=1$ term we  close the integration contour in the upper half-plane and compute it as a sum of residues over single poles. A simplification comes from the fact that due to the term $e^{i\frac{2\pi t}{\epsilon}}$ we need to take into account only the smallest pole $i\frac{\pi}{2}$ since $\epsilon\rightarrow 0$. In the same way one can compute the  $m=-1$ term. Then one finds the CFT normalized ground energy
\begin{equation}
E_0=\lim_{\epsilon, \delta\to 0}\frac{1}{\pi}\left(\mathcal{H}-\mathcal{H}_0\right)=\frac{4}{\delta^2}e^{-\frac{\pi^2}{\epsilon}}\cot{\frac{\pi}{2}(\xi+1)}.
\end{equation}
In the scaling limit it is proportional to $m_1^2$, which is in an agreement with the scaling hypothesis. Finally we arrive to the expected expression for the Restricted Sine-Gordon ground state energy per unit length \cite{AlZam,AlZamTBA,DeVega}
\begin{equation}
    E_0=-\frac{m_1^2}{8\sin{\pi\xi}}\,.
\end{equation}

Let us make a short comment on the technical details. We have to start from the Hamiltonian which provides in the critical point a vanishing ground state energy. The shift of the Hamiltonian provides the constant shift is  magnon energies. This leads to a small modification of the deformation term in the Bethe ansatz equations:
\begin{equation}
    f_{scaling}(u, v)=\exp i\kappa \Big( \left(e(u)-2 \mathcal{H}_0\right)p(v)-\left(e(v)-2 \mathcal{H}_0\right)p(u) \Big).
\end{equation}
Integral equations for the root dencities also change. For example, to get a good scaling limit, we have to modify the equation for a ground state density in the following way
\begin{eqnarray}
    &&\rho^{(\kappa)}_{g}(u)=-\frac{1}{2\epsilon }\frac{d}{du}\ln{\frac{[u+\frac{1}{2}]}{[u-\frac{1}{2}]}}+\frac{1}{2\pi i}\int\limits_{-\frac{i\pi}{2\epsilon}}^{\frac{i\pi}{2\epsilon}}\Bigg(\frac{d}{dv}\ln{\frac{[u-v+1]}{[u-v-1]}}\Bigg)\rho^{(\kappa)}_{g}(v)dv+\cr &&+\frac{i\kappa}{2\epsilon}\Big(\frac{de(u)}{du}\ P^{(\kappa)}_{g}-\left(E^{(\kappa)}_{g}-2 \mathcal{H}_0\right)\frac{dp(u)}{du}  \Big).
    \label{eq_rho_g_scaling}
\end{eqnarray}
What is nice, the whole solution scheme survives. The only difference is in the following change in the answers:
\begin{equation*}
    \mathcal{H}\rightarrow -E_0.
\end{equation*}
As for the excited state and the scattering matrix, we restrict our attention to the case where a couple of breathers has a zero total momentum. Then with the same modification of the Hamiltonian  we obtain the 
energy and the mass of a breather
\begin{eqnarray}
&& 
\mathcal{E}^{(\kappa)}_n(\beta)=m_n^{(\alpha)}\cosh \beta\,.
\cr
&&m^{(\alpha)}_n=\frac{m_n}{1+\alpha E_0}\,,
\end{eqnarray}
These results are in an agreement with the Smirnov-Zamolodchikov's answers for the $T\bar{T}$ perturbation of the restricted sine-Gordon model \cite{TT}. 

In the trigonometric limit the scattering matrix (\ref{SHigherBreathers1}) becomes the breather S-matrix of the sine-Gordon model:
\begin{equation}
\mathcal{S}_{a, b}(\beta_1, \beta_2)= \prod\limits_{l_1=1}^{a}\prod\limits_{l_2=1}^{b}\mathcal{S}_{1,1}\Big(\beta_1-\beta_2-\frac{i\pi\xi}{2}(a-2l_1-b+2l_2)\Big)\,.
\label{SHigherBreathers}
\end{equation}
As before, the scattering matrix of the basic particles is given by the well-known expression \cite{ZZ} 
\begin{equation}
    \mathcal{S}_{1,1}(\beta)=\frac{\tanh{\frac{1}{2}\Big(\beta+i\pi\xi  \Big)}}{\tanh{\frac{1}{2}\Big(\beta-i\pi\xi   \Big)}}\,.
\end{equation}
Finally, the current-current deformed S-matrix which describes the scattering of the breathers $a$ and $b$ has the following form 
\begin{equation}
    \mathcal{S}^{(\alpha)}_{a,b}(\beta)=\mathcal{S}_{a,b}(\beta)\exp\left({\frac{i\alpha}{1+\alpha E_0}m_a m_b\sinh{\beta}}\right)\,.
\end{equation}
We conclude that with some modification the deformation of Bargheer, Beisert, and Loebbert, applied to the scaling Forrester-Baxter model, led to the CDD-factor \cite{CDD} in the scattering matrix as it is in the Smirnov-Zamolodchikov's S-matrix for the $T\bar{T}$-deformed Restricted Sine-Gordon model \cite{TT}. This is our main statement. 

\subsection{Operator approach arguments}
In the previous sections we studied the lattice system directly by solving the Bethe ansatz equations. Now we provide  addition arguments supporting our main result (\ref{finish}). 

We recall the deformation of the lattice Hamiltonian (\ref{mean_derivative}): 
\begin{equation*}
    \frac{d}{d\kappa}{\mathbf H}^{(\kappa)}=\epsilon^{\mu\nu}\int dx \  {\mathbf J}^{(\kappa)}_{1,\mu}(x)\mathbf{J}^{(\kappa)}_{2,\nu}(x),
\end{equation*}
where the conserved currents are related to the deformed theory. After the deformation they remain conserved:
\begin{eqnarray*}
&&
\frac{\partial \mathbf {q}^{(\kappa)}_a}{\partial{t}}+\frac{\partial \mathbf {j}^{(\kappa)}_a}{\partial{x}}=0,  \cr
&&
\frac{\partial \mathbf {q}^{(\kappa)}_a}{\partial{t}}=-i[{\mathbf H}^{(\kappa)},\mathbf {q}^{(\kappa)}_a]\,.
\end{eqnarray*}
Using these equations, one can show that the deformation equation for the Hamiltonian can be rewritten as a sum of two parts (as it was done for the Bose gas in \cite{Jiang}, see also \cite{DeformedS4}). The first one is a commutator with a "bi-local"\ operator $\mathbf X^{(\kappa)}$, while the second takes into account boundary terms: 
\begin{equation}
    \frac{d}{d\kappa} {\mathbf H}^{(\kappa)}=i[{\mathbf X}^{(\kappa)}, {\mathbf H}^{(\kappa)}]+\Big( {\mathbf Q}_1^{(\kappa)} \, \mathbf{j}_2^{(\kappa)}(0)-{\mathbf Q}_2^{(\kappa)} \, \mathbf{j}_1^{(\kappa)}(0) \Big),
    \label{changeH}
\end{equation}
where the periodic boundary conditions $\mathbf{j}_a(L)=\mathbf{j}_a(0)$ are assumed. Here $L$ is the size of the system and ${\mathbf Q}^{(\kappa)}_a$ are the total charges
\begin{equation*}
\mathbf{Q}_a^{(\kappa)}=\int\limits_0^L \mathbf{q}_a^{(\kappa)}(x)dx.
\end{equation*}
The "bi-local"  operator ${\mathbf X}^{(\kappa)}$ is a continuous analogue of the deformation operator (\ref{X}).
\begin{equation*}
    {\mathbf X}^{(\kappa)}=\int\limits_{x<y}dxdy\, \mathbf{q}_1^{(\kappa)}(x)\mathbf{q}_2^{(\kappa)}(y).
\end{equation*}
We claim that the states of the system change under the deformation in the following way:  
\begin{eqnarray}
&&
\ket{\psi^{(\kappa)}}={\mathbf U}^{(\kappa)}\ket{\psi}\,,\cr
&&{\mathbf U}^{(\kappa)}=\exp \Big(-i\int\limits_{0}^{\kappa} {\mathbf X}^{(\lambda)}d\lambda \Big)\,,
    \label{state_def}
\end{eqnarray}
where $\ket{\psi}$ is a state of an undeformed theory. It is possible to find the change of the energy spectrum using the equation (\ref{changeH}):
\begin{eqnarray}
  &&\frac{d}{d\kappa}F^{(\kappa)}= \langle {\psi}^{(\kappa)}|\frac{d}{d\kappa} {\mathbf H}^{(\kappa)}\ket{\psi^{(\kappa)}}=\cr &&= i\ \langle {\psi^{(\kappa)}}|[{\mathbf X}^{(\kappa)}, {\mathbf H}^{(\kappa)}]\ket{\psi^{(\kappa)}}+\langle {\psi^{(\kappa)}}|\Big({\mathbf Q}^{(\kappa)}_1 \mathbf{j}_2^{(\kappa)}(0)-{\mathbf Q}_2^{(\kappa)} \mathbf{j}_1^{(\kappa)}(0)  \Big)\ket{\psi^{(\kappa)}}\,.
  \label{89}
\end{eqnarray}
Since the deformed Hamiltonian acts diagonally on the state $\ket{\psi^{(\kappa)}}$, the first term disappears. The eigenvalue of the second term factorizes into the product of eigenvalues of single operators. There is a simplification if one of the charges has a zero eigenvalue corresponding to the state $\ket{\psi^{(\kappa)}}$, for example $\langle {\psi^{(\kappa)}}|{\mathbf Q}_2^{(\kappa)}\ket{\psi^{(\kappa)}}=0$. In particular, this happened in our case where we considered the ground state and took $\mathbf{Q}_1=\mathbf{H}$ and
  $\mathbf{Q}_2=\mathbf{P}$. In that case (\ref{89}) provides a differential equation for the ground state energy:
\begin{equation}
    \frac{\partial}{\partial{\kappa}}F^{(\kappa)}(L)-F^{(\kappa)}(L)\frac{\partial}{\partial{L}}F^{(\kappa)}(L)=0\,,
    \label{burger}
\end{equation}
which is the well-known Burgers' equation discussed above.
The implicit form of it's solution can be found by the method of characteristics
\begin{equation*}
    F^{(\kappa)}(L)=F_{0}(L+\kappa F^{(\kappa)})\,,
\end{equation*}
where  $F_0(L)$ is the initial condition corresponding to the undeformed case. In our case $F_{0}(L)=L\mathcal{H}$. Then the solution is
\begin{equation}
    F^{(\kappa)}(L)=\frac{L\mathcal{H}}{1-\kappa\mathcal{H}}.
\end{equation}
Exactly this result we obtained using the Bethe ansatz approach in (\ref{change_e_g}) provided the identification $F^{(\kappa)}(L)=LE^{(\kappa)}_g$ .

Let us now discuss general expectations on the structure of the deformed breather S-matrix. We apply arguments from the ref. \cite{LR} which in our situation led to a slightly different final answer. Consider a two-particle state $\ket{\psi_{u,v}}$ given by a superposition of two plane waves:
\begin{equation}
    \ket{\psi_{u,v}}=A(u,v)\ket{\psi_{u<v}}+A(v,u)\ket{\psi_{v<u}}\,.
\label{UndefS}
\end{equation}
The scattering matrix is defined as a ratio of the amplitudes \begin{equation}
    S(u,v)=\frac{A(v,u)}{A(u,v)}\,.
\end{equation}
This is the scattering matrix before the deformation.
Now we turn the current-current deformation on. The wave functions are then changed according to (\ref{state_def}): 
\begin{eqnarray}
&&\frac{d}{d\kappa}\ket{\psi^{(\kappa)}_{u<v}}=-i{\mathbf X}^{(\kappa)}\ket{\psi^{(\kappa)}_{u<v}}\,,\cr
&&\frac{d}{d\kappa}\ket{\psi^{(\kappa)}_{u>v}}=-i{\mathbf X}^{(\kappa)}\ket{\psi^{(\kappa)}_{u>v}}\,.
\end{eqnarray}
Consider, for example, the differential equation for the vector $\ket{\psi^{(\kappa)}_{u<v}}$.  
The operator $\bold{X}^{(\kappa)}$ acts on it as following
\begin{equation}
    {\mathbf X}^{(\kappa)}\ket{\psi^{(\kappa)}_{u<v}}=\Big( f^{(\kappa)}(u)+f^{(\kappa)}(v)+q_1^{(\kappa)}(u)q_2^{(\kappa)}(v) \Big)\ket{\psi^{(\kappa)}_{u<v}},
\end{equation}
where  $q_a^{(\kappa)}(u)$ are the eigenvalues of the corresponding charge densities and $f^{(\kappa)}(u)$ is the result of both charge densities acting on the same particle. 
Then the differential equation for this state reads
\begin{equation}
       \frac{d}{d\kappa}\ket{\psi^{(\kappa)}_{u<v}}=-i\Big(f^{(\kappa)}(u)+f^{(\kappa)}(v)+q_1^{(\kappa)}(u)q_2^{(\kappa)}(v) \Big)\ket{\psi^{(\kappa)}_{u<v}},
\end{equation}
and the similar equation holds for the state $\ket{\psi^{(\kappa)}_{v<u}}$. 
Then the deformation of  S-matrix is given by the following equation: 
\begin{eqnarray*}
\frac{d}{d\kappa}\ln S^{(\kappa)}(u,v)=i\Big(q^{(\kappa)}_1(v)q^{(\kappa)}_2(u)-q^{(\kappa)}_1(u)q^{(\kappa)}_2(v)\Big)\,,
\end{eqnarray*}
where all eigenvalues are related to the deformed system.
We know that in the regime where magnons are physical particles, the eigenvalues of integrals of motion are not deformed. Then it is possible to integrate this equation immediately and get the answer we started from (\ref{magnones}):
\begin{eqnarray*}
S^{(\kappa)}(u,v)=S(u,v)e^{i\kappa(e(v)p(u)-e(u)p(v))}\,,
\end{eqnarray*}
where we already put $\mathbf{Q}_1=\mathbf{H}$ and $\mathbf{Q}_2=\mathbf{P}$.

In the regime III of a Forrester-Baxter model we found that the energy and momentum of a particle do change according to (\ref{PE}). This leads to the following differential equation for the breather S-matrix:
\begin{equation}
    \frac{d}{d\kappa}\ln{\mathcal{S}^{(\kappa)}_{a,b}}(u,v)=i\Big(\mathcal{P}^{(\kappa)}_a(u)\mathcal{E}^{(\kappa)}_b(v)-\mathcal{E}^{(\kappa)}_a(u)\mathcal{P}^{(\kappa)}_b(v)  \Big).
\end{equation}
Using the explicit expressions for the $\kappa$-dependent terms in the right hand side, we easily integrate this equation and arrive to the answer (\ref{finish}):
\begin{equation*}
    \mathcal{S}^{(\kappa)}_{a,b}(u,v)=\mathcal{S}_{a,b}(u,v)e^{ \frac{i\kappa}{1-\kappa \mathcal{H}}\big(\mathcal{P}_{a}(u)\mathcal{E}_{b}(v)-\mathcal{E}_{a}(u)\mathcal{P}_{b}(v)  \big)}.
\end{equation*}
Note that this answer is valid for the magnon S-matrix as well. In that case one should put $\mathcal{H}=0$ and change the breather undeformed energy and momentum by the magnon ones.

\section{Conclusion}
We studied the properties of the ground and excited states of the deformed Forrester-Baxter RSOS models $(2,2s+1)$ in the regime III.

The main aim of our study was to clarify the proposal from the works \cite{TTspin,TTspin1} that the bi-local deformations of  integrable spin chains of Bargheer, Beisert, and Loebbert \cite{LR} are lattice analogues of the $T{\bar T}$ perturbations of  Smirnov and Zamolodchikov \cite{TT}. 

Starting from the deformed Bethe ansatz equations in the thermodynamic limit, we found the density of the ground state  energy. Our answer agrees with the solution of the Burgers' equation \cite{TT} which directly follows from the factorization property of the lattice deformation operator \cite{TTspin}. Using the  deformed Bethe ansatz equations for the excited state, we computed the energies and momenta of  breather-like excitations. 

We checked that the breather-like excitations correspond to the string solutions of the BAE with lengths $l<s$ \cite{Foda, Foda1}. Namely, the expressions for the breather energies and momenta, being computed exactly in the off-critical case, coincide in the trigonometric limit with the corresponding quantities of the  Restricted sine-Gordon model. In the trigonometric limit we obtained the relativistic dispersion  relation and found that the deformation leads to  the rescaling of masses of the breathers. The formulae were derived in all orders of the perturbation theory with the assumption $|\kappa\mathcal{H}|< 1$.  

Using these results, we derived the $\kappa$-deformed breather scattering matrix.  In the trigonometric limit it becomes the S-matrix of the Restricted sine-Gordon model \cite{Freund}  modified by the CDD factor \cite{TT}. As it was expected, the fusion rules in the elliptic situation are the same as in the trigonometric case. Let us note that the results for the pure case $(\kappa=0)$ can be guessed from the previous studies of the eight vertex model as well as from the algebraic arguments related to the Deformed Virasoro algebra \cite{LHP,SL,YP,FR,LPST}. 

Our results, obtained in the Bethe ansatz approach, provide an additional confirmation to the proposal that the bi-local deformations of lattice models are related to the  $T\bar{T}$ perturbations of QFTs. 
However, we would like to point out that there are some subtleties. 

First of all, let us note that our results correspond to the deformation of the lattice model in the form (\ref{changeH}):
\begin{equation*}
    \frac{d}{d\kappa} {\mathbf H}^{(\kappa)}=i[{\mathbf X}^{(\kappa)}, {\mathbf H}^{(\kappa)}]+\Big( {\mathbf Q}_1^{(\kappa)} \, \mathbf{j}_2^{(\kappa)}(0)-{\mathbf Q}_2^{(\kappa)} \, \mathbf{j}_1^{(\kappa)}(0) \Big).
\end{equation*}
The motivation of ref. \cite{LR} was to find new integrable systems for their further application to AdS/CFT correspondence and integrable structures in N=4 super Yang-Mills theory \cite{N4}. The authors considered basically a ferromagnetic regime where the integrals of motion can be chosen to act trivially on the corresponding physical vacuum. In particular, the energy of the ground state is zero. In this case the boundary term in the deformation equation (\ref{changeH}) does not change eigenvalues of conserved charges. However, if one would like to study a paramagnetic sector with breathers-like excitations, this term becomes responsible for the appearance of the non-trivial Burgers' equation (\ref{burger}), which leads to the deformation of masses and to the modification of the CDD factor as was shown in (\ref{finish}). 

Our another observation is related to the subtlety in the definition of the scaling limit. To get the proper CFT-normalized results \cite{TT}, we need to start from the Hamiltonian which in the pure case has a vanishing vacuum expectation value in the critical point.
Subtracting the constant $\mathcal{H}_0$ from the Hamiltonian (\ref{hamiltonian}) slightly modifies the deformed BAE. In the case of the zeroth total momentum such modified construction still leads to the same answers as before, the only change is 
$$
\mathcal{H}\longrightarrow \mathcal{H}-\mathcal{H}_0=E_0\,.
$$
Repeating accurately all computations with the modified Hamiltonian, we found that the current-current deformed lattice models in the scaling limit reproduce  Smirnov-Zamolodchikov's answers for QFTs with the identification
\begin{equation*}
\kappa=-\alpha\,.
\end{equation*}
Without the modification of the Hamiltonian 
one gets an unusual result: the 
quantity $\kappa \mathcal{H}$ becomes infinitely large for a fixed value of $\kappa$ if the lattice step $\delta$ is going to zero. This means that either the original unmodified theory does not have a good continuous limit or our perturbation theory scheme should be improved. 

Nevertheless, we hope that our direct derivations might be useful for a more detailed  understanding of the structure of the $T\bar{T}$ lattice deformation.

There are many open questions which are not considered in the present work. First of all, we could not easily integrate the deformation equation and get an explicit expression for the lattice Hamiltonian, as it was done for simple models of QFT (see for example refs. \cite{CNST,Bonelli,Conti}). This may allow to study interesting problems in deformed lattice models in an analogy with a set of recent QFT results \cite{CNST,Cardy,Conti}. Also it may be interesting to study a wider class of deformations where the perturbation operator is constructed from the integrals of motion with higher spins. As well we expect that the whole scheme can be applicable for regime where the critical point of a RSOS model is described by the parafermionic CFT. From the physical point of view it would be interesting to study the effects of instability and non-locality related to the special form of the solution of the Burgers' equation; these phenomena are actively studied in the $T\bar{T}$ deformed QFTs. Though we did not focused on the sign and the magnitude of the parameter $\kappa$, it is clear that there is a special divergency at the point $\kappa\mathcal{H}=1$. It would be interesting to understand the physics which stands behind this singularity. Finally, after clearing up problems with the ground state, it is natural to try to investigate lattice correlation functions and form factors \cite{AlZam,LHP,MussardoSimon,LPST,TTSinh1} in the current-current perturbed theories. A simple deformation of the formulae which we discussed above gives a hope that these problems are also solvable. 

\section*{Acknowledgements}
We would like to thank H. Babujian, A. Gorsky, M. Lashkevich, A. Litvinov, S. Lukyanov, D. Menskoy, F. Smirnov and E. Sozinov for useful discussions. The work was supported by Russian Foundation of Basic Research under the grant 20-52-05015.

\appendix


\begin{thebibliography}{100}
\bibitem{Baxter:book}
Baxter, R.J.: Exactly Solved Models in Statistical Mechanics. Academic Press, London (1982).

\bibitem{ZZ}
A. B. Zamolodchikov, Al. B. Zamolodchikov, Factorized S Matrices in Two- Dimensions as the Exact Solutions of Certain Relativistic Quantum Field Models. Ann. Phys. 120, 253-291,  (1979).

\bibitem{TT} F. A. Smirnov and A. B. Zamolodchikov, On space of integrable quantum field theories. Nucl. Phys., B 915,  363–383, (2017). arXiv:1608.05499.	

\bibitem{ZamTT} A. Zamolodchikov, Expectation value of composite field $T\bar{T}$ in two-dimensional quantum field theory. hep-th/0401146v1.

\bibitem{AlZamTri} Al. B. Zamolodchikov, From tricritical Ising to critical Ising by thermodynamic Bethe ansatz. Nucl. Phys. B358, 524–546, (1991).

\bibitem{AlZamStair} Al. B. Zamolodchikov, Resonance factorized scattering and roaming trajectories. J. Phys. A 39, 12847 (2006). Preprint ENS-LPS-335, (1991).

\bibitem{Dorey} P. Dorey, F. Ravanini, Generalising the staircase models, Nucl.Phys. B 406, 708-726, (1993).

\bibitem{MussardoSimon} G. Mussardo and P. Simon, Bosonic-type S-matrix, vacuum instability and CDD ambiguities. Nucl. Phys., 
B578, 527–551,  (2000). 

\bibitem{LR} 
T. Bargheer, N. Beisert, and F. Loebbert, Long-range deformations for integrable spin
chains, J. Phys. A, 42, 285205,  (2009). 

\bibitem{TTspin} 
B. Pozsgay, Y. Jiang and G. Takács, TT-deformation and long range spin chains. Journal of High Energy Physics, 92, (2020). arxiv:1911.11118.

\bibitem{TTspin1} E. Marchetto, A. Sfondrini, Z. Yang, TT  Deformations and Integrable Spin Chains. Phys. Rev. Lett. 124,  10, 100601, (2020). arxiv:1911.12315. 

\bibitem{Forrester-Baxter} P. J. Forrester and R. J. Baxter, Further exact solutions of the eight-vertex SOS model and generalizations of the Rogers–Ramanujan identities. J. Statist. Phys. 38, 435, (1985).

\bibitem{BPZ}A.A. Belavin, A.M. Polyakov and A.B. Zamolodchikov, Infinite conformal symmetry in two-dimensional quantum field theory, Nucl. Phys. B241, 333, (1984).

\bibitem{Zam}
A. B. Zamolodchikov, Integrable field theory from conformal field theory. Adv. Stud. Pure Math. 19, 641-674,  (1989). 

\bibitem{Cardy}	John L. Cardy, G. Mussardo, S-matrix of the Yang-Lee edge singularity in two dimensions.  Phys. Lett. B225, (1989). 

\bibitem{AlZam} Al. B. Zamolodchikov, Two-Point correlation function in Scaling Lee-Yang model.
Nucl. Phys. B348, 619-641 (1991).

\bibitem{LeClair} A. LeClair, Restricted Sine-Gordon theory and the minimal conformal series. Phys. Lett. B230, 103-107, (1989).

\bibitem{SmirnovRG}
F. Smirnov, Reductions of the sine-Gordon model as a perturbation of minimal models of conformal field theory. Nucl. Phys. B337, 156–180, (1990).

\bibitem{Freund}  P.G. Freund, T.R. Klassen, E. Melzer, S Matrices for perturbations of certain conformal field theories Phys. Lett. B229, 243 (1989).

\bibitem{BR} V. V. Bazhanov, N. Yu. Reshetikhin, Critical RSOS models and conformal field theory. International Journal of Modern Physics A., 04, 115-142, (1989).

V. V. Bazhanov, N. Yu. Reshetikhin. Scattering amplitudes in off-critical models and RSOS integrable models. Prog. Theor. Phys. Supplement, 102:301– 318, (1990).


\bibitem{Wilson} K.G. Wilson, John B. Kogut, The Renormalization group and the epsilon expansion, Phys. Rept. 12, 75-200, (1974).

\bibitem{AlZamTBA} A. B. Zamolodchikov. Thermodynamic Bethe ansatz in relativistic models. Scaling three state Potts and Lee-Yang models. Nucl. Phys., B342, 695-720 (1990).

\bibitem{CNST} A. Cavaglia, S. Negro, I. M. Szecsenyi, and R. Tateo, TT-deformed 2D Quantum Field Theories. JHEP 10, 112, (2016). arXiv:1608.05534.

\bibitem{DeformedS1} M. Caselle, D. Fioravanti, F. Gliozzi, and R. Tateo, Quantization of the effective string with TBA. JHEP 07, 071,  (2013). arXiv:1305.1278.

\bibitem{DeformedS2} S. Dubovsky, R. Flauger, and V. Gorbenko, Solving the Simplest Theory of Quantum Gravity. JHEP 09, 133,  (2012). arXiv:1205.6805.

\bibitem{DeformedS3} S. Dubovsky, V. Gorbenko, and M. Mirbabayi, Natural Tuning: Towards A Proof of Concept. JHEP 09, 045,  (2013). arXiv:1305.6939.

\bibitem{DeformedS35} S. Dubovsky, V. Gorbenko, M. Mirbabayi, Asymptotic fragility, near AdS2 holography and TT. JHEP 1709, 13,  (2017).

\bibitem{CardyTT} J. Cardy, The $T\bar{T}$ deformation of quantum field theory as random geometry. JHEP 1810, 186 (2018).
		
\bibitem{DeformedS4} J. Kruthoff, O. Parrikar, On the flow of states under $T\bar{T}$. arxiv:2006.03054.


\bibitem{Baxter1972} Baxter, R.J.: Partition function of the eight-vertex lattice model. Ann. Phys. 70, 193–228, (1972).

\bibitem{TahtajianFaddeev} L. A. Takhtadzhan, L. D. Faddeev, The quantum method of inverse problem and the Heisenberg XYZ model. Russ. Math. Surv., 34, 11–68, (1979).

\bibitem{ABF:1984} Andrews, G., Baxter, R. and Forrester, J., Eight-vertex SOS model and generalized Rogers-Ramanujan identities. J. Stat. Phys. 35, 193-266, (1984).

\bibitem{De_Vega}
De Vega, H. J., Face algebras and exact Bete ansatz solutions for SOS models. International Journal of Modern Physics A, 5, 1611-1632, (1990).

\bibitem{LR1} T. Bargheer, N. Beisert, and F. Loebbert, LETTER: Boosting nearest-neighbour to long-range integrable spin chains. Journal of Statistical Mechanics: Theory and Experiment 2008 no. 11, L11001, (2008). arXiv:0807.5081.

\bibitem{LR2}
N. Beisert, L. Fiévet, M. de Leeuw, and F. Loebbert, Integrable deformations of the XXZ spin chain. Journal of Statistical Mechanics: Theory and Experiment 2013, no. 9, 09028,  (2013). arXiv:1308.1584.

\bibitem{Smirnov}  F. A. Smirnov, Form Factors in Completely Integrable Models of Quantum Field Theory. Adv. Series in Math Physics 14, World Scientific, Singapore, (1992).

\bibitem{LHP}	
S. Lukyanov, Y. Pugai, Multi-point local height probabilities in the integrable RSOS model. Nuclear Physics B 473,  631-658, (1996).

\bibitem{SL} S. Lukyanov, A note on the Deformed Virasoro Algebra. Phys. Lett. B367, 121,  (1996). hep-th/9509037.

\bibitem{YP} Y. Pugai, On Vertex Operators and The Normalization of Form Factors, in Statistical Field Theories, edited by A. Cappelli and G. Mussardo, pp. 57–66, Springer Netherlands, Dordrecht, 2002.

\bibitem{FR} D. Fioravanti,  M. Rossi, The elliptic scattering theory of the 1/2-XYZ and higher order Deformed Virasoro Algebras. Ann. Henri Poincaré 7, 1449–1462, (2006).

\bibitem{LPST} M. Lashkevich, Y. Pugai, J. Shiraishi, Y. Tutiya Lattice models, deformed Virasoro algebra and reduction equation. J. Phys. A: Math. Gen., 53 (24), 245202 (2020).

\bibitem{Takahashi_strings} Takahashi, M. and Suzuki, M., One-Dimensional Anisotropic Heisenberg Model at Finite Temperatures.  Progress of Theoretical Physics, 48, No. 6B, p. 2187-2209,  (1972).

\bibitem{KR} Anatol N. Kirillov, N. Yu. Reshetikhin, Exact solution of the integrable XXZ Heisenberg model with arbitrary spin. I. The ground state and the excitation spectrum, J. Phys. A: Math. Gen., v.20, p. 1565, (1987).

\bibitem{Breather_S_matrix} Doikou, A. and  Nepomechie, R., Direct calculation of breather S matrices.  J. Phys. A: Math. Gen. 32,    3663 (1999).

\bibitem{Andrei}
N. Andrei, C. Destri, Dynamical symmetry breaking and fractionization in a new integrable model. Nucl. Phys., B231 (1984)

\bibitem{Korepin} V.E. Korepin, Direct calculation of the S matrix in the massive Thirring model. Teor. Mat. Phys, v 41, 169, (1979).

\bibitem{Hrachik} Hratchya M. Babujian, A.M. Tsvelik, Heisenberg magnet with an arbitrary spin and anisotropic chiral field. Nucl. Phys. B 265, 24-44,  (1986).  
		
\bibitem{Foda} 
A. Berkovich, B. McCoy, A. Schilling, Rogers-Schur-Ramanujan type identities for the M(p,p') minimal models of conformal field theory. Comm. Math. Phys., 191, 325–395 (1998).  arXiv: q-alg/9607020;

\bibitem{Foda1} 	
O. Foda, K. S. M. Lee, Y. Pugai, T. A. Welsh,  Path generating transforms. Contemp. Math. 254,  pp. 157–186, (2000). arXiv:math/9810043

\bibitem{CDD}
L. Castillejo, R. H. Dalitz, F. J. Dyson, Low’s Scattering Equation for the Charged and Neutral Scalar Theories. Phys. Rev. 101, 453, (1956).

\bibitem{JMS} M. Jimbo, T. Miwa, F. Smirnov, Fermionic structure in the sine-Gordon model: Form factors and null-vectors. Nucl. Phys. B 852, 390-440,  (2011).

\bibitem{DelfinoNikoli1} G. Delfino, G. Niccoli, Matrix elements of the operator TT-bar in integrable quantum field theory. Nucl.Phys. B707, 381-404,  (2005). hep-th/0407142.

\bibitem{DelfinoNikoli2} G. Delfino, G. Niccoli, The Composite operator T anti-T in sinh-Gordon and a series of massive minimal models. JHEP 0605, 035,  (2006). hep-th/060222.

\bibitem{DelfinoNikoli3} G. Delfino and G. Niccoli, Form factors of descendant operators in the massive Lee–Yang model. J. Stat. Mech. 0504, P04004, (2005). hep-th/0501173.

\bibitem{TTSinh} 
M. Lashkevich, Y. Pugai, Note on four-particle form factors of operators $T_{2n}T_{-2n}$  in sinh-Gordon model. J. Phys. A49, 305401, (2016).

\bibitem{TTSinh1}
M. Lashkevich, Y. Pugai, The complex sinh-Gordon model: form factors of descendant operators and current-current perturbations. JHEP 01, 071,  (2019).

\bibitem{Jiang}
Y. Jiang, $T\bar{T}$-deformed 1d Bose gas.
arxiv:2011.00637.

\bibitem{DeVega} Destri, C. and de Vega, H.: New exact result in affine Toda field theories: Free energy and wave functional renormalization. Nucl. Phys., B358, 251-294 (1991).

\bibitem{LZ}
S. Lukyanov, A. Zamolodchikov,
Exact expectation values of local fields in quantum sine-Gordon model. Nucl.Phys. B493 571-587, (1997). 

\bibitem{Bonelli} G. Bonelli, N. Doroud, M. Zhu,
$T\bar{T}$-deformations in closed form. JHEP 1806, 149, (2018). 

\bibitem{Conti} 
R. Conti, S. Negro, R. Tateo, The $T\bar{T}$ perturbation and its geometric interpretation. JHEP 02, 085  (2019).


\bibitem{N4} N. Beisert et al., Review of AdS/CFT Integrability: An Overview.
Lett. Math. Phys. 99, 3 (2012), arxiv:1012.3982.

\end{thebibliography}
\end{document}